\documentclass[12pt]{article}
\pdfoutput=1

\usepackage{amsmath,amssymb,amsfonts,graphicx,amsfonts}
\usepackage{textcomp,mathcomp}
\usepackage{color}
\usepackage[hidelinks]{hyperref}
\usepackage{ascmac}
\usepackage{physics}

\allowdisplaybreaks[4]

\date{}
\begin{document}
\begin{flushright}
\today\\
\end{flushright}

\vspace{0.1cm}

\begin{center}
  {\Large Toward QCD on Quantum Computer: Orbifold Lattice Approach}\\

\end{center}
\vspace{0.1cm}
\vspace{0.1cm}
\begin{center}

Georg Bergner$^{a}$, Masanori Hanada$^{b}$, Enrico Rinaldi$^{c}$, Andreas Sch\"{a}fer$^d$\\


\end{center}
\vspace{0.3cm}

\begin{center}

{\small

$^a$Institute for Theoretical Physics, University of Jena\\
Max-Wien-Platz 1, D-07743 Jena, Germany\\
\vspace{3mm}
$^b$School of Mathematical Sciences, Queen Mary University of London\\
Mile End Road, London, E1 4NS, United Kingdom\\
\vspace{1mm}
$^{b}$qBraid Co., Harper Court 5235, Chicago, IL 60615, USA\\
\vspace{3mm}
$^c$Quantinuum K.K., Otemachi Financial City Grand Cube 3F, 1-9-2 Otemachi,\\
Chiyoda-ku, Tokyo, Japan\\
\vspace{1mm}
$^c$Center for Quantum Computing (RQC), RIKEN\\
Wako, Saitama 351-0198, Japan\\
\vspace{1mm}
$^c$Theoretical  Quantum  Physics  Laboratory, RIKEN,
Wako, Saitama 351-0198, Japan\\
\vspace{1mm}
$^c$Interdisciplinary Theoretical and Mathematical Sciences (iTHEMS) Program, RIKEN\\
Wako, Saitama 351-0198, Japan\\
\vspace{3mm}
$^d$Institute of Theoretical Physics, University of Regensburg\\
Universit\"{a}tsstrasse 31, D-93053 Regensburg, Germany\\


}
\end{center}

\vspace{1.5cm}

\begin{center}
  {\bf Abstract}
\end{center}

We propose an orbifold lattice formulation of QCD suitable for quantum simulations. We show explicitly how to encode gauge degrees of freedom into qubits using noncompact variables, and how to write down a simple truncated Hamiltonian in the coordinate basis. We show that SU(3) gauge group variables and quarks in the fundamental representation can be implemented straightforwardly on qubits, for arbitrary truncation of the gauge manifold.

\newpage
\tableofcontents

\section{Introduction}
\hspace{0.51cm}
Quantum Chromodynamics (QCD), and more generally gauge theories with various matter contents, are important targets for quantum simulations~\cite{Bauer:2022hpo,Bauer:2023qgm,Wiese:2013uua,Zohar:2015hwa,Dalmonte:2016alw,Banuls:2019bmf,Aidelsburger:2021mia,Zohar:2021nyc,Klco:2021lap}.
However, to accomplish a viable quantum simulation of real gauge theories (in contrast to toy models of various sort) is a most ambitious task with many aspects.
For example, one would need to start by formulating a representation of the Hamiltonian on logical (or physical) qubits; this includes finding the optimal truncation scheme for gauge fields and a hardware realization of matter fields.
Then one would need to design quantum algorithms to prepare initial states, perform time evolution, and realize the required measurements to access physical observables.
In view of these many requirements it is highly advisable to study a broad selection of Hamiltonian lattice formulations in parallel, rather than just the Kogut-Susskind Hamiltonian, and the more diverse ones, the better.

In this paper, we take the first steps to design the formulation of lattice gauge theories based on orbifold lattices.
We try to highlight what kind of advantages the orbifold lattice formulation has for quantum simulations of lattice gauge theories, compared to the Kogut-Susskind formulation.

The orbifold lattice construction was developed originally to realize exact supersymmetry on a lattice~\cite{Kaplan:2002wv,Cohen:2003xe,Cohen:2003qw,Kaplan:2005ta}.
A big difference from other lattice formulations such as Wilson's plaquette action~\cite{Wilson:1974sk} or Kogut-Susskind Hamiltonian~\cite{Kogut:1974ag} (see e.g., Ref.~\cite{Zohar:2014qma} for a modern survey) is the use of noncompact variables with flat measure to describe gauge field. 
Because of this property, the Hamiltonian formulation is very simple and hence the orbifold lattice approach offers a good alternative to the Kogut-Susskind formulation that can make quantum simulation of Yang-Mills theory rather straightforward~\cite{Buser:2020cvn}.\footnote{
See e.g., Refs.~\cite{Hayata:2023bgh,Gustafson:2022xdt} for attempts to simplify the Kogut-Susskind Hamiltonian. 
Another potentially useful approach, which might work for certain theories, is to find a simpler theory such as a spin model that shows the same universal behaviors as gauge theory at long distance~\cite{Alexandru:2023qzd,Bhattacharya:2020gpm}. 
}

In this paper, we go beyond the proposal made in Ref.~\cite{Buser:2020cvn} and address a few points crucial for quantum simulation of QCD using the orbifold lattice approach. 
Sec.~\ref{sec:review} gives a review of the orbifold lattice formulation of Yang-Mills theory. 
Then, in Sec.~\ref{sec:coordinate_basis}, we give a truncation scheme that results in a finite-dimensional Hamiltonian suitable for quantum simulations. 
A naive construction using the orbifold lattice approach leads to a representation of U($N$) gauge groups rather than SU($N$) ones. 
We propose a resolution to this issue in Sec.~\ref{sec:U-to-SU}. 
In Sec.~\ref{sec:fermion}, we discuss how fermions in the fundamental representation can be introduced. 
Specifically, we discuss naive fermions and Wilson fermions. 
We do not find any particular obstruction such that the simplicity of the orbifold lattice approach survives with fundamental fermions. 
Note that the complications people encountered in the past are associated with supersymmetry, which is not an issue for the application to QCD. 
In summary:
\begin{itemize}
\item
We provide a simple encoding of the truncated Hamiltonian suitable for quantum simulation on qubit-based digital quantum computers. 

\item
We show that fermions in the fundamental representation can be incorporated in a straightforward manner.

\item
We describe how to realize the symmetries of the SU(3) gauge group.  

\end{itemize}
With this work we hope that the orbifold lattice formulation will be considered by the readers as a new starting point for quantum simulations of lattice gauge theories. We leave the realization of a full quantum simulation algorithm for SU(3) lattice gauge theory in the orbifold lattice formulation to future works.

\section{Review of orbifold lattice}\label{sec:review}
\hspace{0.51cm}
In this section, we review the orbifold lattice formulation of pure U($N$) Yang-Mills theory in $3+1$ dimensions. 
The Lagrangian and Hamiltonian formulations are given in Sec.~\ref{sec:Lagrangian} and Sec.~\ref{sec:operator-formalism}, respectively. 
This section is largely a repetition of Ref.~\cite{Buser:2020cvn} and it is included to make this paper self-contained. 

The name `orbifold' lattice comes from the use of the orbifold projection of the matrix model in the original papers~\cite{Kaplan:2002wv,Cohen:2003xe,Cohen:2003qw,Kaplan:2005ta}, where the orbifold projection played the crucial role to construct supersymmetric lattice gauge theory. 
In this paper, our motivation to use the orbifold lattice is different: we want to use it as an alternative to the Kogut-Susskind Hamiltonian formulation, and we do not consider supersymmetry. 
Therefore, our main message can be understood without knowing anything about the orbifold projection. 
For the interested readers, we review the orbifold projection in Appendix~\ref{sec:orbifold-projection}.

\subsection{Lagrangian formulation}\label{sec:Lagrangian}
\hspace{0.51cm}
We discuss spatial lattices with continuous time in Sec.~\ref{sec:Lagrangian-spatial-lattice}, and Euclidean spacetime lattices in Sec.~\ref{sec:spacetime-lattice}. 
\subsubsection{Spatial lattices with continuous time}\label{sec:Lagrangian-spatial-lattice}
Let $\vec{n}=(n_x,n_y,n_z)=(n_1,n_2,n_3)$ be the label of spatial lattice sites, where $n_1,n_2,n_3=1,2,\cdots,L$.
The dynamical degrees of freedom are three $N\times N$ complex matrices $Z_{j,\vec{n}}$ ($j=1,2,3$) living on the links connecting $\vec{n}$ and $\vec{n}+\hat{j}$.
They are related to the unitary link variables $U_{j,\vec{n}}$ as 
\begin{align}
Z_{j,\vec{n}}
=
\frac{1}{g_{4d}}\sqrt{\frac{a}{2}}W_{j,\vec{n}}U_{j,\vec{n}}\, , 
\label{relation-x-vs-U}
\end{align}
where $W_{j,\vec{n}}$ are Hermitian matrices with non-negative eigenvalues, $a$ is the lattice spacing, and $g_{\rm 4d}$ is the bare Yang-Mills coupling constant. $W_{j,\vec{n}}$ and $U_{j,\vec{n}}$ are related to the Hermitian scalar fields $s_{j,\vec{n}}$ and the Hermitian gauge fields $A_{j,\vec{n}}$ by 
\begin{align}
W_{j,\vec{n}}
=
e^{ag_{\rm 4d}s_{j,\vec{n}}}\, , 
\qquad
U_{j,\vec{n}}
=
e^{iag_{\rm 4d}A_{j,\vec{n}}}\, .
\label{eq:continuum-fields}
\end{align}
We use a bar to denote Hermitian conjugation: $\bar{Z}_{j,\vec{n}}=Z_{j,\vec{n}}^\dagger$.
We also have the U($N$) gauge field $A_{t,\vec{n}}$ living on each site $\vec{n}$. 
The Lagrangian is 
\begin{align}
L
&=
\sum_{\vec{n}}
{\rm Tr}\Biggl(
\sum_{j=1}^3|D_tZ_{j,\vec{n}}|^2
-
\frac{g_{\rm 4d}^2}{2a^3}\left|\sum_{j=1}^3
\left(
Z_{j,\vec{n}} \bar{Z}_{j,\vec{n}} -\bar{Z}_{j,\vec{n}-\hat{j}}Z_{j,\vec{n}-\hat{j}}
\right)
\right|^2 
\nonumber\\
&\qquad\qquad\qquad
-
\frac{2g_{\rm 4d}^2}{a^3}\sum_{j<k}
\left|
Z_{j,\vec{n}}Z_{k,\vec{n}+\hat{j}}
-
Z_{k,\vec{n}}Z_{j,\vec{n}+\hat{k}}
\right|^2
 \Biggl)\, . 
\label{eq:lattice-action}
\end{align}
Here, our notation is ${\rm Tr}|M|^2={\rm Tr}(MM^\dagger)$ for any matrix $M$.
The covariant derivative is defined by 
\begin{align}
D_tZ_{j,\vec{n}}
&=
\partial_tZ_{j,\vec{n}}
+
ig_{\rm 4d}A_{t,\vec{n}}Z_{j,\vec{n}}
-
ig_{\rm 4d}Z_{j,\vec{n}}A_{t,\vec{n}+\hat{j}}\, . 
\end{align}
A local U($N$) transformation is given by 
\begin{align}
Z_{j,\vec{n}}
\to
\Omega_{\vec{n}}Z_{j,\vec{n}}\Omega^{-1}_{\vec{n}+\hat{j}}\, , 
\qquad
A_{t,\vec{n}}
\to
\Omega_{\vec{n}}A_{t,\vec{n}}\Omega^{-1}_{\vec{n}}
+
ig_{\rm 4d}^{-1}\Omega_{\vec{n}}\partial_t\Omega^{-1}_{\vec{n}}\, , 
\end{align}
where $\Omega_{\vec{n}}$ is an $N\times N$ unitary matrix at each lattice site $\vec{n}$. 
In terms of $W_{j,\vec{n}}$ and $U_{j,\vec{n}}$, this transformation is
\begin{align}
W_{j,\vec{n}}
\to
\Omega_{\vec{n}}W_{j,\vec{n}}\Omega^{-1}_{\vec{n}}\, , 
\qquad
U_{j,\vec{n}}
\to
\Omega_{\vec{n}}U_{j,\vec{n}}\Omega^{-1}_{\vec{n}+\hat{j}}\, . 
\end{align}

If we assume that $W_j$ and $U_j$ are close to the identity, we can write the Lagrangian as 
\begin{align}
L
=
\int d^3x{\rm Tr}
\left(
-
\frac{1}{4}F_{\mu\nu}^2
-
\frac{1}{2}(D_\mu s_I)^2
+
\frac{g_{\rm 4d}^2}{4}[s_I,s_J]^2
\right)
\end{align}
up to $O(a)$ corrections. 
Thus, it is straightforward to obtain a lattice regularization of $(3+1)$-d YM theory coupled to three scalar fields $s_{j=1,2,3}$. 
More precisely, lattice gauge theory for Yang-Mills theory coupled to adjoint scalars can be obtained if 
\begin{align}
Z_{j,\vec{n}}\bar{Z}_{j,\vec{n}}
=
\frac{a}{2g_{4d}^2}\cdot
W_{j,\vec{n}}^2
\simeq
\frac{a}{2g_{4d}^2}\cdot\textbf{1}_N
\label{eq:how-to-choose-background-2}
\end{align}
is satisfied. 

The problem is that $W_{j,\vec{n}}$ does not stay close to the identity unless we add mass to the scalars. 
To cure this problem, we add
\begin{align}
\Delta L
&\equiv
-\frac{m^2g_{\rm 4d}^2}{2a}\sum_{\vec{n}}\sum_{j=1}^3
{\rm Tr}
\left| Z_{j,\vec{n}}\bar{Z}_{j,\vec{n}} -\frac{a}{2g_{\rm 4d}^2}\right|^2
\nonumber\\
&\quad
-
\frac{N\mu^2g_{\rm 4d}^2}{2a}\sum_{\vec{n}}\sum_{j=1}^3
\left| \frac{1}{N}{\rm Tr}(Z_{j,\vec{n}}\bar{Z}_{j,\vec{n}}) -\frac{a}{2g_{\rm 4d}^2}\right|^2 
\end{align}
to the Lagrangian. 
This term is equivalent to the scalar mass term in the continuum theory:
\begin{align}
\Delta L
&=
-\frac{m^2}{2}\int d^3x{\rm Tr}
\left(
s_1^2+s_2^2+s_3^2
\right) 
\nonumber\\
&\quad
-\frac{\mu^2}{2N}\int d^3x
\left(
({\rm Tr}s_1)^2+({\rm Tr}s_2)^2+({\rm Tr}s_3)^2
\right) 
\end{align}
up to $O(a)$ corrections. 
With this term, the assumption \eqref{eq:how-to-choose-background-2} is justified. 
Note that the second term is the so-called U(1) scalar mass~\cite{Hanada:2010qg} that plays a crucial role to stabilize the lattice structure. 
With radiative corrections, although the SU($N$) part tends to clump up, the U(1) part can run away and the orbifold lattice can be destroyed unless a large mass is added. 
We can tame this instability by taking $\mu$ large, without affecting the dynamics of the SU($N$) part. 
Furthermore, by taking $m^2$ very large, we can eliminate the scalars from low-energy dynamics.\footnote{
In the original papers on the orbifold construction~\cite{Kaplan:2002wv,Cohen:2003xe,Cohen:2003qw,Kaplan:2005ta}, such a mass term was problematic because the goal was to realize supersymmetry on the lattice and the scalar mass breaks supersymmetry. For our current purpose, there is no harm because we are interested neither in supersymmetry nor scalars. 
} 

The gauge-invariant path-integral measure is the flat measure on ${\mathbb R}^{2N^2}$, 
\begin{align}
\int dx_{\mu,\vec{n}}d\bar{x}_{\mu,\vec{n}}
=
\int_{-\infty}^\infty dx^{\rm (R)}_{\mu,\vec{n}}
\int_{-\infty}^\infty dx^{\rm (I)}_{\mu,\vec{n}}\, ,
\end{align}
where $x^{\rm (R)}_{\mu,\vec{n}}$ and $x^{\rm (I)}_{\mu,\vec{n}}$ are the real and imaginary parts of $x_{\mu,\vec{n}}$.
Therefore, the path integral is defined using the flat measure in the same way as in a system of particles in flat space.
This makes the operator formalism very simple, as we will see in Sec.~\ref{sec:operator-formalism}. 

\subsubsection{Spacetime lattice}\label{sec:spacetime-lattice}
\hspace{0.51cm}
Although our primary motivation is quantum simulation, a spacetime lattice can still play an important role because we can use Euclidean lattice Monte Carlo simulations on classical devices to study many features of the theory. For example, we can determine the optimal lattice parameters for quantum simulations. 

In the original papers on the spacetime orbifold lattice~\cite{Cohen:2003xe,Cohen:2003qw,Kaplan:2005ta}, the motivation was supersymmetry. Therefore, orbifold projections to a zero-dimensional supersymmetric matrix models (the IKKT matrix model~\cite{Ishibashi:1996xs} and its variants with smaller supersymmetry) were performed and supersymmetric lattice actions were obtained. In this approach, the temporal link variable was also complex.  

Because we are not interested in supersymmetry, we can follow a different path. We can simply discretize the time direction, introducing unitary link variables and replacing $|D_tZ_j|^2$ with a plaquette, with temporal lattice spacing $a_{\rm t}$. The Euclidean action obtained this way is
\begin{align}
S_{\rm 4d\ lattice}
&=
\sum_{\vec{n}}
{\rm Tr}\Biggl(
\frac{1}{a_{\rm t}}\sum_{j=1}^3 
\left|
U_{t,\vec{n}}Z_{j,\vec{n}+\hat{t}}
-
Z_{j,\vec{n}}U_{t,\vec{n}+\hat{j}}
\right|^2
\nonumber\\
&\qquad\qquad
+
\frac{g_{\rm 4d}^2a_{\rm t}}{2a^3}\left|\sum_{j=1}^3
\left(
Z_{j,\vec{n}} \bar{Z}_{j,\vec{n}} -\bar{Z}_{j,\vec{n}-\hat{j}}Z_{j,\vec{n}-\hat{j}}
\right)
\right|^2 
\nonumber\\
&\qquad\qquad
+
\frac{2g_{\rm 4d}^2a_{\rm t}}{a^3}\sum_{j<k}
\left|
Z_{j,\vec{n}}Z_{k,\vec{n}+\hat{j}}
-
Z_{k,\vec{n}}Z_{j,\vec{n}+\hat{k}}
\right|^2
 \Biggl)
 \ +\ \Delta S_{\rm 4d\ lattice}\, , 
\label{eq:4d_lattice-action}
\end{align}
\begin{align}
\Delta S_{\rm 4d\ lattice}
&\equiv
\frac{m^2g_{\rm 4d}^2a_{\rm t}}{2a}\sum_{\vec{n}}\sum_{j=1}^3
{\rm Tr}
\left| Z_{j,\vec{n}}\bar{Z}_{j,\vec{n}} -\frac{a}{2g_{\rm 4d}^2}\right|^2
\nonumber\\
&\quad
+
\frac{N\mu^2g_{\rm 4d}^2a_{\rm t}}{2a}\sum_{\vec{n}}\sum_{j=1}^3
\left| \frac{1}{N}{\rm Tr}(Z_{j,\vec{n}}\bar{Z}_{j,\vec{n}}) -\frac{a}{2g_{\rm 4d}^2}\right|^2\, . 
\end{align}
Note that $\vec{n}$ contains four integers because now we consider a four-dimensional spacetime lattice. 
When the continuum limit along the time direction ($a_{\rm t}\to 0$ fixing the physical size of temporal direction) is taken without modifying the spatial lattice spacing $a$ and the spatial lattice size, we can get the Euclidean version of the theory discussed in Sec.~\ref{sec:Lagrangian-spatial-lattice}.
\subsubsection{\texorpdfstring{U($N$)}{U(N)} vs \texorpdfstring{SU($N$)}{SU(N)}}
In the formulation discussed above, the gauge symmetry was U($N$) and not SU($N$). 
This is not problematic for pure Yang-Mills, because the U(1) part decouples trivially. 
However, when fundamental fermions are introduced, U(1) does not decouple.

In the Hamiltonian formulation (which we will discuss shortly), the Hamiltonian has U($N$) invariance, rather than SU($N$). Whether we treat the U(1) part as gauge symmetry is a matter of taste, but certainly, we should remove those degrees of freedom that do not exist in QCD. 

In the path integral formulation, we can choose between the gauge symmetry U($N$) and SU($N$) by allowing or forbidding $A_t$ to have a U(1) part. Specifically, we can choose if the Gauss law is imposed on U($N$) or SU($N$). Keeping the U(1) part and having the U($N$) gauge symmetry makes more sense, because otherwise Lorentz symmetry is lost even if scalar fields decouple. (Hence we assumed this above.) Either way, the fact that we have extra degrees of freedom that do not exist in QCD is problematic. 

Therefore, we need to find a way to remove the U(1) part, if we want to study real-world QCD with SU(3) gauge group. 
We will discuss this in Sec.~\ref{sec:U-to-SU}. 

\subsection{Hamiltonian formalism}\label{sec:operator-formalism}
\hspace{0.51cm}
For the operator formulation, we set the temporal gauge field to zero, $A_t=0$, and replace the coordinate $Z_{j,\vec{n}}$ and its conjugate momentum $P_{j,\vec{n}}=\partial_tZ_{j,\vec{n}}$ by operators that satisfy the canonical commutation relation. The integration over $A_t$ in the path integral formalism is equivalent to imposing the gauge-singlet constraint on the states in Hilbert space. (See e.g., Refs.~\cite{Hanada:2020uvt,Rinaldi:2021jbg} for details.)

Specifically, we use the \textit{extended Hilbert space} $\mathcal{H}_{\rm ext}$ defined by using the coordinate eigenstates $\ket{Z}$:
\begin{align}
\mathcal{H}_{\rm ext}
=
\mathrm{Span}\left\{
\ket{Z}\ ; \hat{Z}_{j,\vec{n}}\ket{Z}=Z_{j,\vec{n}}\ket{Z}
\right\}\, .
\end{align}
More precisely, we consider the states $\ket{\Phi}=\int dZ\Phi(Z)\ket{Z}$ with the square-integrable wave function $\Phi(Z)$. 

The Hamiltonian can be written in terms of the link variables $Z_{j,\vec{n}}$, and their canonical conjugates $P_{j,\vec{n}}$ as follows: 
\begin{align}
\hat{H}
&=
\sum_{\vec{n}}
{\rm Tr}\Biggl(
\sum_{j=1}^3 \hat{P}_{j,\vec{n}}\hat{\bar{P}}_{j,\vec{n}}
+
\frac{g_{\rm 4d}^2}{2a^3}\left|\sum_{j=1}^3
\left(
\hat{Z}_{j,\vec{n}} \hat{\bar{Z}}_{j,\vec{n}} -\hat{\bar{Z}}_{j,\vec{n}-\hat{j}}\hat{Z}_{j,\vec{n}-\hat{j}}
\right)
\right|^2 
\nonumber\\
&\qquad\qquad\qquad
+
\frac{2g_{\rm 4d}^2}{a^3}\sum_{j<k}
\left|
\hat{Z}_{j,\vec{n}}\hat{Z}_{k,\vec{n}+\hat{j}}
-
\hat{Z}_{k,\vec{n}}\hat{Z}_{j,\vec{n}+\hat{k}}
\right|^2
 \Biggl)
 + 
 \Delta\hat{H}\, . 
\label{eq:Hamiltonian}
\end{align}
where
\begin{align}
\Delta\hat{H}
&\equiv
\frac{m^2g_{\rm 4d}^2}{2a}\sum_{\vec{n}}\sum_{j=1}^3{\rm Tr}
\left|
\hat{Z}_{j,\vec{n}}
\hat{\bar{Z}}_{j,\vec{n}}
-
\frac{a}{2g_{\rm 4d}^2}
\right|^2
\nonumber\\
&\quad
+
\frac{N\mu^2g_{\rm 4d}^2}{2a}\sum_{\vec{n}}\sum_{j=1}^3
\left| \frac{1}{N}{\rm Tr}(\hat{Z}_{j,\vec{n}}\hat{\bar{Z}}_{j,\vec{n}}) -\frac{a}{2g_{\rm 4d}^2}\right|^2\, . 
\label{Hamiltonian-mass}
\end{align}

The commutation relations are
\begin{align}
[\hat{Z}_{j,\vec{n},pq},\hat{\bar{P}}_{k\vec{n}',rs}]=i\delta_{jk}\delta_{\vec{n}\vec{n}'}\delta_{ps}\delta_{qr},
\label{eq:commutation-relation-KKU}
\end{align}
and
\begin{align}
[\hat{Z},\hat{P}]
=
[\hat{\bar{Z}},\hat{\bar{P}}]
=
[\hat{Z},\hat{Z}]
=
[\hat{Z},\hat{\bar{Z}}]
=
[\hat{\bar{Z}},\hat{\bar{Z}}]
=
[\hat{P},\hat{P}]
=
[\hat{P},\hat{\bar{P}}]
=
[\hat{\bar{P}},\hat{\bar{P}}]
=
0\, . 
\end{align}
Using $^{\rm (R)}$ and $^{\rm (I)}$ to denote real (Hermitian) and imaginary (anti-Hermitian) parts of $\hat{Z}$ as $\hat{Z}=\frac{\hat{Z}^{\rm (R)}+i\hat{Z}^{\rm (I)}}{\sqrt{2}}$,
we obtain 
\begin{align}
[\hat{Z}^{\rm (R)}_{j,\vec{n},pq},\hat{P}^{\rm (R)}_{k,\vec{n}',rs}]
=
[\hat{Z}^{\rm (I)}_{j,\vec{n},pq},\hat{P}^{\rm (I)}_{k,\vec{n}',rs}]
=
i\delta_{jk}\delta_{\vec{n}\vec{n}'}\delta_{ps}\delta_{qr}\, . 
\label{eq:commutation-relation-KKU-Hermitian}
\end{align}
We can introduce adjoint index $\alpha$ such that
\begin{align}
\hat{Z}^{\rm (R)}_{pq}
=
\sum_{\alpha=1}^{N^2}
\hat{Z}^{\rm (R)}_\alpha\tau^\alpha_{pq}\, , 
\qquad
\hat{Z}^{\rm (I)}_{pq}
=
\sum_{\alpha=1}^{N^2}
\hat{Z}^{\rm (I)}_\alpha\tau^\alpha_{pq}\, , 
\end{align}
where $\tau^\alpha$ are generators of U($N$) that satisfy $\mathrm{Tr}(\tau^\alpha\tau^\beta)=\delta^{\alpha\beta}$ and $\sum_{\alpha}\tau^\alpha_{pq}\tau^\alpha_{rs}=\delta_{ps}\delta_{qr}$. Then, $\left(\hat{Z}^{\rm (R)}_\alpha\right)^\dagger=\hat{Z}^{\rm (R)}_\alpha$ and $\left(\hat{Z}^{\rm (I)}_\alpha\right)^\dagger=\hat{Z}^{\rm (I)}_\alpha$, and
\begin{align}
[\hat{Z}^{\rm (R)}_{j,\vec{n},\alpha},\hat{P}^{\rm (R)}_{k,\vec{n}',\beta}]
=
[\hat{Z}^{\rm (I)}_{j,\vec{n},\alpha},\hat{P}^{\rm (I)}_{k,\vec{n}',\beta}]
=
i\delta_{jk}\delta_{\vec{n}\vec{n}'}\delta_{\alpha\beta}\, . 
\end{align}
Therefore, $\hat{Z}^{\rm (R)}_{j,\vec{n},\alpha}$ and $\hat{P}^{\rm (R)}_{j,\vec{n},\alpha}$, and also $\hat{Z}^{\rm (I)}_{j,\vec{n},\alpha}$ and $\hat{P}^{\rm (I)}_{j,\vec{n},\alpha}$, behave the same way as usual coordinate $\hat{x}$ and momentum $\hat{p}$ that satisfy $[\hat{x},\hat{p}]=i$. 

The generators of a local U($N$) transformation at spatial lattice site $\vec{n}$ are 
\begin{align}
\hat{G}_{\vec{n},pq}
\equiv
i\sum_{j=1}^3
\left(
-
\hat{Z}_{j,\vec{n}}\hat{\bar{P}}_{j,\vec{n}}
+
\hat{P}_{j,\vec{n}}\hat{\bar{Z}}_{j,\vec{n}}
-
\hat{\bar{Z}}_{j,\vec{n}-\hat{j}}\hat{P}_{j,\vec{n}-\hat{j}}
+
\hat{\bar{P}}_{j,\vec{n}-\hat{j}}\hat{Z}_{j,\vec{n}-\hat{j}}
\right)_{pq}. 
\label{eq:gauge-generators}
\end{align}
Indeed, we can confirm that they act properly on the links attached to the site $\vec{n}$:
\begin{align}
\left[
\sum_{r,s}
\epsilon^{rs}\hat{G}_{\vec{n},sr},\hat{Z}_{\mu,\vec{n},pq}
\right]
=
-(\epsilon\hat{Z}_{\mu,\vec{n}})_{pq}, 
\qquad
\left[
\sum_{r,s}
\epsilon^{rs}\hat{G}_{\vec{n},sr},\hat{Z}_{\mu,\vec{n}-\hat{\mu},pq}
\right]
=
(\hat{Z}_{\mu,\vec{n}-\hat{\mu}}\epsilon)_{pq}. 
\end{align}
The Hamiltonian is invariant under local U($N$) transformations:
\begin{align}
[\hat{G}_{\vec{n}},\hat{H}]=0\, . 
\end{align}

Corresponding to the integration over $A_t$ in the path integral formalism, the extended Hilbert space $\mathcal{H}_{\rm ext}$ is projected to the gauge-invariant Hilbert space $\mathcal{H}_{\rm inv}$ consisting of U($N$)-singlet states. We can incorporate this gauge condition by using only the singlet states as initial condition, or by interpreting the non-singlet states related by the local U($N$) transformations as the same state. 

\subsection{Plaquette model}\label{sec:plaquette}
\hspace{0.51cm}
Some readers may find the orbifold lattice discussed above rather complicated and wonder why we do not simply use the plaquette term consisting of $Z$s, i.e., 
\begin{align}
\hat{H}_{\rm plaquette}
&=
\sum_{\vec{n}}
{\rm Tr}\left(
\sum_{j=1}^3 \hat{P}_{j,\vec{n}}\hat{\bar{P}}_{j,\vec{n}}
-
\frac{2g_{\rm 4d}^2}{a^3}\sum_{j\neq k}
\hat{Z}_{j,\vec{n}}\hat{Z}_{k,\vec{n}+\hat{j}}\hat{\bar{Z}}_{j,\vec{n}+\hat{k}}\hat{\bar{Z}}_{k,\vec{n}}
\right)
 + 
 \Delta\hat{H}\, ,  
\label{eq:plaq-Hamiltonian}
\end{align}
where $\Delta\hat{H}$ is the same as \eqref{Hamiltonian-mass}.
Probably this option can also work, because we add a large mass to scalars and make them decouple. 
However, a tricky issue associated with this plaquette Hamiltonian is that this is not bounded from below and the scalar fields $s_{1,2,3}$ are pushed to infinity.
To compensate this, we need to choose the mass parameter in $\Delta\hat{H}$ large (specifically, $m\gtrsim\frac{1}{a}$). 
Such a large parameter could introduce technical complications in actual simulations.
\section{Qubit realization in the coordinate basis}\label{sec:coordinate_basis}
\hspace{0.51cm}
To put an orbifold lattice on a qubit-based quantum computer, we need to truncate the Hilbert space to finite dimensions.  
In the past, the truncation in the harmonic oscillator basis was studied in some detail~\cite{Buser:2020cvn}. 
The coordinate basis was mentioned only briefly in an appendix of Ref.~\cite{Buser:2020cvn}. 
In this section, we show that the truncation in the coordinate basis~\cite{Jordan:2012xnu,Jordan:2011ci,Klco:2018zqz} can make the truncated Hamiltonian very simple. 

Let $\{\ket{\vec{x}}\}$ be the coordinate basis for a particle in flat space that satisfies
\begin{align}
\hat{\vec{x}}\ket{\vec{x}}
=
\vec{x}\ket{\vec{x}}\, , 
\end{align}
where $\vec{x}\equiv (x_1,x_2,\cdots)$ stands for real and imaginary parts of $Z_{j,\vec{n}}$. 
This state $\ket{\vec{x}}$ is related to the coordinate eigenstate of each boson as 
\begin{align}
\ket{\vec{x}}
=
\otimes_i\ket{x_i}\, .
\end{align}
Below, we focus on the truncation of the single-boson Hilbert space. We can easily obtain the truncated Hilbert space of orbifold lattice theory by taking a tensor product. 

We introduce a cutoff for the value of $x$, 
\begin{align}
-R\le x\le R\, , 
\end{align}
and discretize $x$ by introducing $\Lambda$ points, 
\begin{align}
x_n
=
-R+n\delta_x\, ,
\qquad
\delta_x=\frac{2R}{\Lambda-1}\, , 
\qquad
n=0,1,\cdots,\Lambda-1\, .   
\end{align}
The truncation parameters $\Lambda$, $\delta_x$ and $R$ should be sent to $\infty$, 0 and $\infty$, respectively.  
By using $\ket{n}$ to denote $\ket{x_n}$, we can write the operator $\hat{x}$ as
\begin{align}
\hat{x}
=
\sum_{n=0}^{\Lambda-1}
x_{n}
\ket{n}\bra{n}
=
-R\cdot\textbf{1}
+\delta_x\cdot\hat{n}\, , 
\end{align}
where
\begin{align}
\hat{n}
\equiv
n
\ket{n}\bra{n}
\end{align}
is the number operator. 

To rewrite it as a sum of Pauli operators, let us use the binary from, 
\begin{align}
\ket{n} =\ket{b_1}\ket{b_2}\cdots\ket{b_K}\, , 
\qquad
b_i=0\ \mathrm{or}\ 1\, , 
\qquad
n=b_1+2b_2\cdots+2^{K-1}b_K\, . 
\end{align}
Here, $\Lambda=2^K$. 
Then, the number operator can be written by using Pauli $\sigma_z$ gates, 
\begin{align}
\hat{n}
=
\frac{\hat{\sigma}_{z,1}+1}{2}
+
2\cdot\frac{\hat{\sigma}_{z,2}+1}{2}
+
\cdots
+
2^{K-1}\cdot\frac{\hat{\sigma}_{z,K}+1}{2}\, , 
\end{align}
where $\hat{\sigma}_{z,i}$ is the Pauli $\hat{\sigma}_{z}$ operator acting on $\ket{b_i}$. 

\subsection{Interaction part}
Because the interaction terms consist of four coordinate operators, it is written as a sum of products of at most four $\hat{\sigma}_{z}$s.
When we consider that interaction in the exponential of the Hamiltonian (i.e. when doing time evolution), the unitary operator can be realized with a quantum circuit that has single-qubit gates and two-qubit gates following standard decomposition methods~\cite{Cowtan:2019loc}.
Efficient algorithms for the decomposition of the exponential of multi-$\hat{\sigma}_{z}$ operators into quantum circuits are standard in most quantum compilation frameworks and are the subject of intense research to improve the performance of quantum hardware~\cite{Sriluckshmy:2023leq,Algaba:2023enr}.
An example of the implementation of $e^{i \alpha \hat{\sigma}_{z}\hat{\sigma}_{z}\hat{\sigma}_{z}\hat{\sigma}_{z}}$ using $R_{Z Z}=e^{-i \frac{\theta}{2} \hat{\sigma}_{z} \hat{\sigma}_{z}}$ quantum gates that are native to trapped ion devices was presented in Ref.~\cite{Shaydulin:2023fpr}.
A decomposition using CNOT gates on superconducting qubits was recently shown in Ref.~\cite{Farrell:2024fit}.

If we use the Fock-space truncation, the interaction terms become more complicated~\cite{Buser:2020cvn}. Because the interaction part contains more terms than the momentum part, it is reasonable to adopt a truncation scheme which makes the interaction part simpler.

\subsection{Momentum part}

Next, we consider the momentum part. We consider $\hat{p}^2$ acting on the single-boson Hilbert space, because ${\rm Tr}(\hat{P}_{j,\vec{n}}\hat{\bar{P}}_{j,\vec{n}})=\frac{1}{2}\sum_{\alpha=1}^{N^2}\left((\hat{P}^{\rm (R)}_{j,\vec{n},\alpha})^2+(\hat{P}^{\rm (I)}_{j,\vec{n},\alpha})^2\right)$ is simply a sum of  such terms. 

To simplify the expressions, we impose the periodic boundary condition $\ket{\Lambda}=\ket{0}$. 
This is possible because, when the boundary condition matters, the truncated model is not a good approximation of the original model.  
In this case, the shift operator $\hat{S}\equiv\sum_n\ket{n+1}\bra{n}$ is identified with $e^{i\delta_X\hat{p}}$. Therefore, $\hat{p}=\frac{\hat{S}^{1/2}-\hat{S}^{-1/2}}{i\delta_X}$ up to the corrections of order $\delta_X$. 
Then, $\hat{p}^2=\frac{2\hat{I}-\hat{S}-\hat{S}^{-1}}{\delta_X^2}$, and hence
\begin{align}
\hat{p}^2
=
\frac{2\hat{I}-\hat{S}-\hat{S}^{-1}}{\delta_X^2}
=
\frac{1}{\delta_X^2}
\sum_{n=0}^{\Lambda-1}
\left\{
2\ket{n}\bra{n}
-
\ket{n+1}\bra{n}
-
\ket{n}\bra{n+1}
\right\}\, . 
\label{p^2-coordinate-basis}
\end{align}
Because $\sum_n\ket{n}\bra{n}$ is the identity, the nontrivial parts of $\hat{p}^2$ are $\hat{S}=\sum_n\ket{n+1}\bra{n}$ and $\hat{S}^{-1}$. Basically, we only need addition and subtraction of 1, which are rather elementary operations.  

We show three ways to handle this $\hat{p}^2$ below.

\subsubsection*{Quantum Fourier Transform}
An obvious option is to use momentum eigenstate to handle the momentum part of the Hamiltonian. 
Quantum Fourier transform\footnote{We do not use `QFT' to abbreviate Quantum Fourier Transform because it is already assigned to Quantum Field Theory.} allows us to do this. For each bosonic degree of freedom represented by $K$ qubits ($\Lambda=2^K$), the number of gates needed for quantum Fourier transform is of order $K^2$. To each momentum eigenstate $\hat{p}$, $\hat{S}$ acts as multiplication by $e^{ip\delta_X}$.
The cost of the inverse Quantum Fourier Transform that is needed to go back to the coordinate basis is again of order $K^2$. 

See e.g., Refs.~\cite{Alam:2021uuq,Gustafson:2022xdt} for a treatment of the momentum part via quantum Fourier transform. 

\subsubsection*{NOT, CNOT and CCNOT with auxiliary qubits}
Another simple option is to replace logic gates on classical computer by corresponding quantum gates. 
This can be done by using $K-1$ auxiliary qubits set to $\ket{0}$ (which can be reduced to two qubits if we reuse the same qubits after resetting them), one NOT (=$\hat{\sigma}_x$), $K$ CNOT, and $K-2$ CCNOT gates. The concrete construction is explained in Appendix~\ref{sec:plus_1}. 

\subsubsection*{Pauli-string expansion}
Yet another way to realize $S$ is to use Pauli strings. For each $n$, we can write $n$ and $n+1$ in a binary form as $n=\sum 2^{l-1}b_l$ and $n+1=\sum 2^{l-1}b'_l$, and then, express $\ket{b'_l}\bra{b_l}$ in terms of Pauli matrices, e.g., $\ket{0}\bra{1}=\frac{\hat{\sigma}_x+i\hat{\sigma}_y}{2}$. Combining it with $\ket{n+1}\bra{n}=\otimes_l\ket{b'_l}\bra{b_l}$, we get the Pauli-string expression of $\ket{n+1}\bra{n}$. 
\subsection{Generators of \texorpdfstring{SU($N$)}{SU(N)} transformation}
To construct a truncated version of SU($N$) generators \eqref{eq:gauge-generators}, we need $\hat{p}$ rather than $\hat{p}^2$. Associated with this, a slight complication is that $\frac{\hat{S}^{1/2}-\hat{S}^{-1/2}}{i\delta_X}$ takes a messy form in the coordinate basis. Instead, we can use $\frac{\hat{S}-1}{i\delta_X}$ to approximate $\hat{p}$. Though $\frac{\hat{S}-1}{i\delta_X}$ is not Hermitian, we can keep $\hat{G}$ Hermitian if we use $\frac{\hat{S}-1}{i\delta_X}$ for $\hat{P}$ and $(\frac{\hat{S}-\textbf{1}}{i\delta_X})^\dagger=\frac{\textbf{1}-\hat{S}^{-1}}{i\delta_X}$ for $\hat{\bar{P}}$.

\section{Projecting \texorpdfstring{U($N$)}{U(N)} to \texorpdfstring{SU($N$)}{SU(N)}}\label{sec:U-to-SU}
\hspace{0.51cm}
In this section, we discuss how the U(1) part in U($N$) can be projected out so that we can get SU($N$) theory. 

The simplest option is to do nothing. If U(1) is not asymptotically free, it decouple automatically at low energy, so we can just forget about it. 

If we want to remove the U(1) part explicitly, we can add terms that force $\det Z_{j,\vec{n}}$ to be 1, such as $\sum_j|\det Z_{j,\vec{n}} - c^N|^2$ times a large number, where $c=\frac{1}{\sqrt{2}ag_{1d}}$. For SU(2), it is not more complicated than the mass term for scalars; $\det Z_{j,\vec{n}}$ consists of 2 quadratic terms, and $|\det Z_{j,\vec{n}} - c^N|^2$ contains 4 complex quartic terms. For SU(3), $\det Z_{j,\vec{n}}$ consists of 6 cubic terms, and hence, $|\det Z_{j,\vec{n}} - c^N|^2$ contains 36 complex sextic terms. Quartic terms or sextic terms can be expressed by using tensor products of at most four or six $\hat{\sigma}_z$s, respectively. 

We can also reduce the number of terms by taking $|{\rm Im}\det Z_{j,\vec{n}}|^2$ instead of $|\det Z_{j,\vec{n}}- c^N|^2$.~\footnote{
Setting ${\rm Im}\det Z_{j,\vec{n}}$ to zero restricts the U(1) part to $\pm 1$. When studying the Hamiltonian time evolution, if the phase is initially chosen to be $+1$, it remains $+1$.
} Let's see how this second option works. 

\textbf{SU(2):}
The determinant is $\det Z_j = Z_{j,11}Z_{j,22}-Z_{j,12}Z_{j,21}$. In terms of real (Hermitian) and imaginary (anti-Hermitian) parts defined by $Z=\frac{Z^{\rm (R)}+iZ^{\rm (I)}}{\sqrt{2}}$, 
\begin{align}
|\mathrm{Im}\det Z_j|^2 = \frac{1}{4}\left(Z_{j,11}^{\rm (R)}Z_{j,22}^{\rm (I)}+Z_{j,11}^{\rm (I)}Z_{j,22}^{\rm (R)}-Z_{j,12}^{\rm (R)}Z_{j,21}^{\rm (I)}-Z_{j,12}^{\rm (I)}Z_{j,21}^{\rm (R)}\right)^2\, .
\end{align} 
We can introduce adjoint index $\alpha$ by using $\tau^{1,2,3}=\frac{\sigma^{1,2,3}}{\sqrt{2}}$ and $\tau^4=\frac{\textbf{1}}{\sqrt{2}}$. Then, we can rewrite the above expression by using self-adjoint variables $Z_{j,\alpha}^{\rm (R)}$ and $Z_{j,\alpha}^{\rm (I)}$:
\begin{align}
|\mathrm{Im}\det Z_j|^2 = \frac{1}{4}\left(
Z_{j,\alpha=1}^{\rm (R)}Z_{j,\alpha=1}^{\rm (I)}
+
Z_{j,\alpha=2}^{\rm (R)}Z_{j,\alpha=2}^{\rm (I)}
+
Z_{j,\alpha=3}^{\rm (R)}Z_{j,\alpha=3}^{\rm (I)}
-
Z_{j,\alpha=4}^{\rm (R)}Z_{j,\alpha=4}^{\rm (I)}
\right)^2\, .
\end{align} 
The right-hand side contains 10 quartic terms of self-adjoint coordinate variables. 

\textbf{SU(3):} The determinant $\det Z_j$ contains 6 complex terms. Expressed by $Z^{\rm (R)}$ and $Z^{\rm (I)}$, there are $6\times2^3=48$ terms. Half of them are in $\mathrm{Im}\det Z_j$. Therefore, $|\mathrm{Im}\det Z_j|^2$ contains $24+\frac{24\cdot 23}{2}=300$ terms.

\section{Fermions}\label{sec:fermion}
\hspace{0.51cm}
In this section, we discuss how Dirac fermions in the fundamental representation can be incorporated into the orbifold lattice construction. 
The Lagrangian of the continuum theory in Minkowski signature is 
\begin{align}
i\int d^4x \bar{\psi}(\gamma^\mu D_\mu - m)\psi\, , 
\end{align}
where gamma matrices $\gamma^\mu$ ($\mu=0,1,2,3$) are $4\times 4$ matrices that satisfy
\begin{align}
\{\gamma^\mu,\gamma^\nu\}=2\eta^{\mu\nu}\, ,
\qquad
(\gamma^\mu)^\dagger=\gamma_\mu\, , 
\qquad
\gamma^0\gamma_\mu\gamma^0=\gamma^\mu\, , 
\end{align}
$\eta=\mathrm{diag}(-,+,+,+)$, and $\bar{\psi}$ is defined by  
\begin{align}
\bar{\psi}=\psi^\dagger\gamma^0\, . 
\label{eq:psi-bar-psi-dagger}
\end{align}
For the Euclidean signature, $\gamma^0D_0$ is replaced with $\gamma^4D_4$, where $\gamma^4=i\gamma^0$. 
Specifically, we discuss how the naive and Wilson fermions have to be modified for noncompact gauge variables instead of unitary variables.  
Roughly speaking, we simply replace $U_i$ with $\sqrt{\frac{2g_{\rm 4d}^2}{a}}Z_i =W_iU_i$. When scalars are heavy, $W_j$ approaches $\textbf{1}$, and hence there is no difference.

Below, we start with the Lagrangian formulation on spacetime lattices.  Then, we study the Lagrangian and Hamiltonian formulations for the spatial lattices with continuous time. 

\subsection{Lagrangian formulation}
\hspace{0.51cm}

\subsubsection{Naive fermions on a spacetime lattice}
\hspace{0.51cm}
Let us consider the naive fermion first. We replace spatial unitary link variables 
in the naive fermion action with unitary link variables $U_{j=1,2,3}$ with $\sqrt{\frac{2g_{\rm 4d}^2}{a}}Z_j = W_jU_j$, which can be expanded as $W_jU_j=e^{ag_{\rm 4d}s_j}U_j=U_j+ag_{\rm 4d}s_jU_j+O(a^2)$. Then, 
\begin{align}
S_{\rm naive}
&=
ia^3a_t\sum_{\vec{n}}\Bigl\{
-\frac{1}{2a_{\rm t}}
\left(
\bar{\psi}_{\vec{n}}\gamma^4 U_{4,\vec{n}}\psi_{\vec{n}+\hat{4}}
-
\bar{\psi}_{\vec{n}+\hat{4}}\gamma^4 U^\dagger_{4,\vec{n}}\psi_{\vec{n}}
\right)
\nonumber\\
&\quad
-\sqrt{\frac{2g_{\rm 4d}^2}{a}}\cdot
\frac{1}{2a}
\sum_{j=1}^3
\left(
\bar{\psi}_{\vec{n}}\gamma^j Z_{j,\vec{n}}\psi_{\vec{n}+\hat{j}}
-
\bar{\psi}_{\vec{n}+\hat{j}}\gamma^j \bar{Z}_{j,\vec{n}}\psi_{\vec{n}}
\right)
+
m\bar{\psi}_{\vec{n}}\psi_{\vec{n}}
\Bigl\}\, . 
\end{align}
In terms of continuum fields, the additional terms associated with the replacement of $U_j$ with $\sqrt{\frac{2g_{\rm 4d}^2}{a}}Z_j = W_jU_j$ are
\begin{align}
\frac{iag_{\rm 4d}}{2}
\int d^4x\sum_{j=1}^3
\left(
2\bar{\psi}\gamma^j s_{j}D_j\psi
+
\bar{\psi}\gamma^j (D_js_{j})\psi
\right)
+
\cdots\, .
\end{align}
We have an additional interaction term with heavy scalars that could break Lorentz symmetry if they did not decouple. We can take the mass of the scalar as large as the cutoff scale. Furthermore, interactions in the UV are suppressed by an overall factor $a^1$. 
Thus, such terms are unlikely to be harmful. 
\subsubsection{Wilson fermions on spacetime lattice}
\hspace{0.51cm}
The Wilson term is 
\begin{align}
S_{\rm Wilson}
&=
ia^3a_{\rm t}\Bigl\{
\frac{1}{2a}\sum_{j=1}^3\sum_{\vec{n}}\left(
\left(\bar{\psi}_{\vec{n}}-\bar{\psi}_{\vec{n}+\hat{j}}U^\dagger_{j,\vec{n}}\right)
\left(\psi_{\vec{n}}-U_{j,\vec{n}}\psi_{\vec{n}+\hat{j}}\right)
\right)
\nonumber\\
&\qquad
+
\frac{1}{2a_{\rm t}}\sum_{\vec{n}}\left(
\left(\bar{\psi}_{\vec{n}}-\bar{\psi}_{\vec{n}+\hat{4}}U^\dagger_{4,\vec{n}}\right)
\left(\psi_{\vec{n}}-U_{4,\vec{n}}\psi_{\vec{n}+\hat{4}}\right)
\right)
\Bigl\}\, . 
\end{align}

Again, we replace $U_{j=1,2,3}$ with $\sqrt{\frac{2g_{\rm 4d}^2}{a}}Z_j=W_jU_j$, which is written as $W_jU_j=e^{ag_{\rm 4d}s_j}U_j=U_j+ag_{\rm 4d}s_jU_i+O(a^2)$, and we obtain the orbifold-lattice version of the Wilson term as
\begin{align}
S_{\rm Wilson}
&=
ia^3a_{\rm t}\Bigl\{
\frac{1}{2a}\sum_{j=1}^3\sum_{\vec{n}}\left(
\left(\bar{\psi}_{\vec{n}}-\sqrt{\frac{2g_{\rm 4d}^2}{a}}\bar{\psi}_{\vec{n}+\hat{j}}\bar{Z}_{j,\vec{n}}\right)
\left(\psi_{\vec{n}}-\sqrt{\frac{2g_{\rm 4d}^2}{a}}Z_{j,\vec{n}}\psi_{\vec{n}+\hat{j}}\right)
\right)
\nonumber\\
&\qquad
+
\frac{1}{2a_{\rm t}}\sum_{\vec{n}}\left(
\left(\bar{\psi}_{\vec{n}}-\bar{\psi}_{\vec{n}+\hat{4}}U^\dagger_{4,\vec{n}}\right)
\left(\psi_{\vec{n}}-U_{4,\vec{n}}\psi_{\vec{n}+\hat{4}}\right)
\right)
\Bigl\}\, . 
\label{orbifold-wilson-term}
\end{align}
Then, the additional term on top of the original Wilson term is 
\begin{align}
\frac{iag_{\rm 4d}}{2}\sum_{j=1}^3\int d^4x\bar{\psi}(D_js_j)\psi+\cdots\, , 
\end{align}
which does not look particularly harmful. 

In \eqref{orbifold-wilson-term}, there is a term proportional to $\bar{\psi}_{\vec{n}+\hat{j}}\bar{Z}_{j,\vec{n}}Z_{j,\vec{n}}\psi_{\vec{n}+\hat{j}}$. There are options to change this to $\bar{\psi}_{\vec{n}}Z_{j,\vec{n}}\bar{Z}_{j,\vec{n}}\psi_{\vec{n}}$ or $\frac{1}{2}\bar{\psi}_{\vec{n}+\hat{j}}\bar{Z}_{j,\vec{n}}Z_{j,\vec{n}}\psi_{\vec{n}+\hat{j}}+\frac{1}{2}\bar{\psi}_{\vec{n}}Z_{j,\vec{n}}\bar{Z}_{j,\vec{n}}\psi_{\vec{n}}$. The same is the case for Lagrangian and Hamiltonian on spatial lattices. The last option appears better because of the spatial reflection symmetry. 
\subsubsection{Spatial lattice}
\hspace{0.51cm}
Next, we consider the spatial lattice with Minkowski signature. 
Again, we simply replace $U_{j=1,2,3}$ with $\sqrt{\frac{2g_{\rm 4d}^2}{a}}Z_j=W_jU_j$. 
The naive fermion can be generalized to the orbifold lattice as 
\begin{align}
L_{\rm naive}
=
ia^3\sum_{\vec{n}}\left\{
\psi^\dagger_{\vec{n}}D_t\psi_{\vec{n}}
+
\frac{1}{2a}\sqrt{\frac{2g_{\rm 4d}^2}{a}}
\sum_{j=1}^3
\left(
\bar{\psi}_{\vec{n}}\gamma^j Z_{j,\vec{n}}\psi_{\vec{n}+\hat{j}}
-
\bar{\psi}_{\vec{n}+\hat{j}}\gamma^j \bar{Z}_{j,\vec{n}}\psi_{\vec{n}}
\right)
-
m\bar{\psi}_{\vec{n}}\psi_{\vec{n}}
\right\}\, . 
\label{naive_fermion_spatial_lattice}
\end{align}
Note that $-\bar{\psi}\gamma^0D_0\psi=\psi^\dagger D_t\psi$. 
The Wilson term can be generalized as 
\begin{align}
L_{\rm Wilson}
=
-
a^3\cdot\frac{i}{2a}\sum_{j=1}^3\sum_{\vec{n}}
\left(\bar{\psi}_{\vec{n}}-\sqrt{\frac{2g_{\rm 4d}^2}{a}}\bar{\psi}_{\vec{n}+\hat{j}}\bar{Z}_{j,\vec{n}}\right)
\left(\psi_{\vec{n}}-\sqrt{\frac{2g_{\rm 4d}^2}{a}}Z_{j,\vec{n}}\psi_{\vec{n}+\hat{j}}\right)\, . 
\label{Wilson_fermion_spatial_lattice}
\end{align}

\subsection{Hamiltonian formulation}
\hspace{0.51cm}

The naive fermion Hamiltonian can be generalized to the orbifold lattice as 
\begin{align}
\hat{H}_{\rm naive}
=
ia^3\sum_{\vec{n}}\left\{
-
\frac{1}{2a}\sqrt{\frac{2g_{\rm 4d}^2}{a}}
\sum_{j=1}^3
\left(
\hat{\bar{\psi}}_{\vec{n}}\gamma^j \hat{Z}_{j,\vec{n}}\hat{\psi}_{\vec{n}+\hat{j}}
-
\hat{\bar{\psi}}_{\vec{n}+\hat{j}}\gamma^j \hat{\bar{Z}}_{j,\vec{n}}\hat{\psi}_{\vec{n}}
\right)
+
m\hat{\bar{\psi}}_{\vec{n}}\hat{\psi}_{\vec{n}}
\right\}\, . 
\end{align}
The Wilson term can be generalized as 
\begin{align}
\hat{H}_{\rm Wilson}
=
a^3\cdot\frac{i}{2a}\sum_{j=1}^3\sum_{\vec{n}}
\left(\hat{\bar{\psi}}_{\vec{n}}-\sqrt{\frac{2g_{\rm 4d}^2}{a}}\hat{\bar{\psi}}_{\vec{n}+\hat{j}}\hat{\bar{Z}}_{j,\vec{n}}\right)
\left(\hat{\psi}_{\vec{n}}-\sqrt{\frac{2g_{\rm 4d}^2}{a}}\hat{Z}_{j,\vec{n}}\hat{\psi}_{\vec{n}+\hat{j}}\right)\, . 
\end{align}

As mentioned before, there are options to replace $\hat{\bar{\psi}}_{\vec{n}+\hat{j}}\hat{\bar{Z}}_{j,\vec{n}}\hat{Z}_{j,\vec{n}}\hat{\psi}_{\vec{n}+\hat{j}}$ with $\hat{\bar{\psi}}_{\vec{n}}\hat{Z}_{j,\vec{n}}\hat{\bar{Z}}_{j,\vec{n}}\hat{\psi}_{\vec{n}}$ or $\frac{1}{2}\hat{\bar{\psi}}_{\vec{n}+\hat{j}}\hat{\bar{Z}}_{j,\vec{n}}\hat{Z}_{j,\vec{n}}\hat{\psi}_{\vec{n}+\hat{j}}+\frac{1}{2}\hat{\bar{\psi}}_{\vec{n}}\hat{Z}_{j,\vec{n}}\hat{\bar{Z}}_{j,\vec{n}}\hat{\psi}_{\vec{n}}$. The last option appears better because of the spatial reflection symmetry. 

The canonical anti-commutation relations are
\begin{align}
\{
\hat{\psi}_{\vec{n},\alpha},\hat{\psi}^\dagger_{\vec{n}',\alpha'}
\}
=
\delta_{\vec{n}\vec{n}'}
\delta_{\alpha\alpha'}\, , 
\qquad
\{
\hat{\psi}_{\vec{n},\alpha},\hat{\psi}_{\vec{n}',\alpha'}
\}
=
\{
\hat{\psi}^\dagger_{\vec{n},\alpha},\hat{\psi}^\dagger_{\vec{n}',\alpha'}
\}
=
0\, . 
\end{align}
Note that $\bar{\psi}$ and $\psi^\dagger$ are related by \eqref{eq:psi-bar-psi-dagger}.

\section{Future directions}
\hspace{0.51cm}
In this paper, we showed that the orbifold lattice offers a convenient tool for quantum simulation of QCD.
Specifically, we saw that the use of noncompact variables and the coordinate basis truncation of the Hilbert space enable us to simplify the realization on digital quantum computers compared to approaches based on unitary variables.
Potential advantages include the following:
\begin{itemize}
\item
Unlike in the Kogut-Susskind formulation, we do not use any details of the gauge group representations, which will change for different SU($N$) groups. 
We use the same truncation scheme regardless of the gauge group of the theory.
The explicit form of the truncated Hamiltonian is given without using group-theoretic factors, such as Clebsh-Gordan coefficients.
In the literal sense, \textbf{we can write down the Hamiltonian explicitly for any gauge group, lattice size, and truncation levels}.
Therefore, we can systematically study the effect of truncation, complexity of simulation algorithms, and so on, in a straightforward way.
\item
The coordinate basis may have advantage over the momentum basis (electric basis) for the simulation of the continuum limit, because the weak-coupling vacuum takes a simpler form in the coordinate basis.
The strongly-coupled vacuum, which can be described directly in the electric basis, might be separated by a phase transition from the weak-coupling regime that is relevant for the continuum limit. 
The Kogut-Susskind formulation is not amenable to a straightforward design of a truncation scheme in the coordinate basis; see Refs.~\cite{Jakobs:2023lpp,Garofalo:2023zkd,Romiti:2023hbd,DAndrea:2023qnr} for recent attempts.
In the orbifold lattice formulation we have provided a direct truncation scheme in the coordinate basis.
\item
The use of noncompact variables makes quantum simulations with continuous variables, such as photonic quantum computing, an appealing option as well. 
\end{itemize}

We believe that we provided the readers with strong enough motivations for studying the orbifold lattice approach to QCD. 
Admittedly, however, our discussions focused only on the formulation of the Hamiltonian and lacked a detailed investigation of simulation algorithms, truncation effects, and so forth.
There are many directions that must be explored by using pen, paper, classical computer, and quantum computer.
Here is an incomplete list:
\begin{itemize}
\item
To describe QCD as efficiently as possible, we need a precise understanding of physics on an orbifold lattice. 
How should we choose the parameters? Should we take the bare scalar mass as large as the cutoff scale,\footnote{Note that the U(1) mass $\mu$ must be large in order to stabilize the orbifold lattice, but the mass of the SU($N$) part $m$ does not have to be so large because it receives a large radiative correction.} or should we take it somewhere between the QCD scale and the cutoff scale? How do the scalars affect low-energy dynamics as function of their bare mass? 
Note also that, when the U(1) scalar mass is large but finite, `lattice spacing' could change dynamically. It is important to determine such a `renormalized' lattice spacing. 
All this can be studied analytically or numerically, without using quantum devices.
\item
In this paper, we discussed only the naive and Wilson fermions. We should also study other fermions and understand how we can realize chiral symmetry. 
In this context, it is important to note that exact symmetries on the orbifold lattice control the radiative corrections so that we can avoid fine tunings when we take the continuum limit. We will discuss more on this issue in a separate publication.
\item
In this paper, we did not discuss how fermions can be encoded on qubits. It is important to find an efficient scheme that is optimal for the orbifold lattice approach.

\item
The coordinate basis may enable us to prepare the initial state for quantum simulation adiabatically starting with the weak-coupling limit, and avoiding the bulk phase transition. In this context, it is also important to study and compare the phase diagrams of orbifold and Kogut-Susskind Hamiltonian and understand the pros and cons of each formulation. 

\item
The method introduced in Ref.~\cite{Hanada:2022pps} can be used to estimate the truncation effect of the orbifold lattice in the coordinate-basis truncation scheme. We might be able to give precise resource estimate which is intractable for exact diagonalization of Hamiltonians. A nontrivial issue is a potential interplay between the digitization step size $\delta_x$ and scalar mass. Smaller $\delta_x$ would be needed for larger scalar mass.

\item
The use of noncompact variables may allow us to use analog quantum simulators. 

\item
Quantum simulation on an orbifold lattice and for a matrix model could be technically almost identical. Hence, simulation of a matrix model with small matrix size could be the best first step toward quantum simulation of QCD. Obviously, this is also an important step toward the study of quantum gravity via holography. Note also that Hamiltonians of the SYK model~\cite{Sachdev:2015efa,Maldacena:2016hyu} and randomly-coupled spin models~\cite{Hanada:2023rkf,Swingle:2023nvv}, which serve as good toy models for holography, share essential features with the interaction term of a matrix model and orbifold lattice gauge theory. 
In this sense, quantum simulation of QCD and \textit{quantum gravity in the lab}~\cite{Danshita:2016xbo,Brown:2019hmk,Gharibyan:2020bab,Maldacena:2023acv} are closely related. 
Note also that the motivation of the original papers on the orbifold lattice~\cite{Kaplan:2002wv,Cohen:2003xe,Cohen:2003qw,Kaplan:2005ta} was a study of quantum gravity through lattice simulation of dual supersymmetric gauge theory~\cite{KU-private-communication}.

\end{itemize}
\begin{center}
\Large{\textbf{Acknowledgement}}
\end{center}
The authors would like to thank Hidehiko Shimada and Pavlos Vranas for discussions. 
M.~H. thanks his STFC consolidated grant ST/Z001072/1. 
M.~H. and E.~R. thank the Royal Society International Exchanges award IEC/R3/213026.
G.~B.\ is funded by the Deutsche Forschungsgemeinschaft (DFG) under Grant No.~432299911 and 431842497.

\appendix
\section{Orbifold projection from a matrix model}\label{sec:orbifold-projection}
\hspace{0.51cm}
In this section, we review how the orbifold lattice can be obtained from a matrix model through orbifold projection. 
As emphasized in the main text, our main result can be understood without referring to this construction. 
We added this appendix just to please those readers who want to better understand the historical background.

Below, we first show how the Lagrangian \eqref{eq:lattice-action} is obtained from a matrix model following Refs.~\cite{Kaplan:2002wv,Buser:2020cvn}, and then show how the fermion part is obtained. 

\begin{figure}[htbp]
\begin{center}
\scalebox{0.4}{
\includegraphics{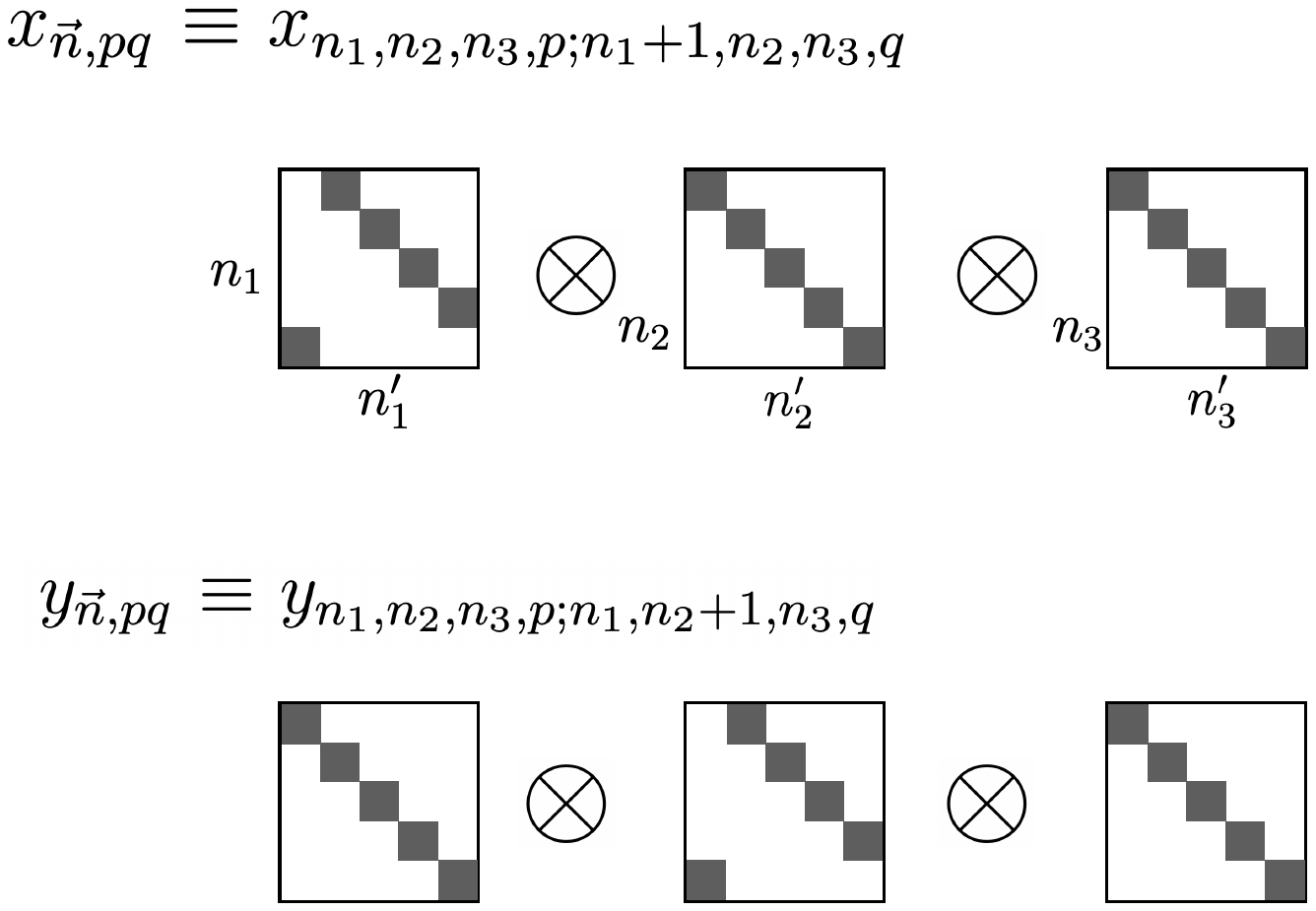}}
\scalebox{0.4}{
\includegraphics{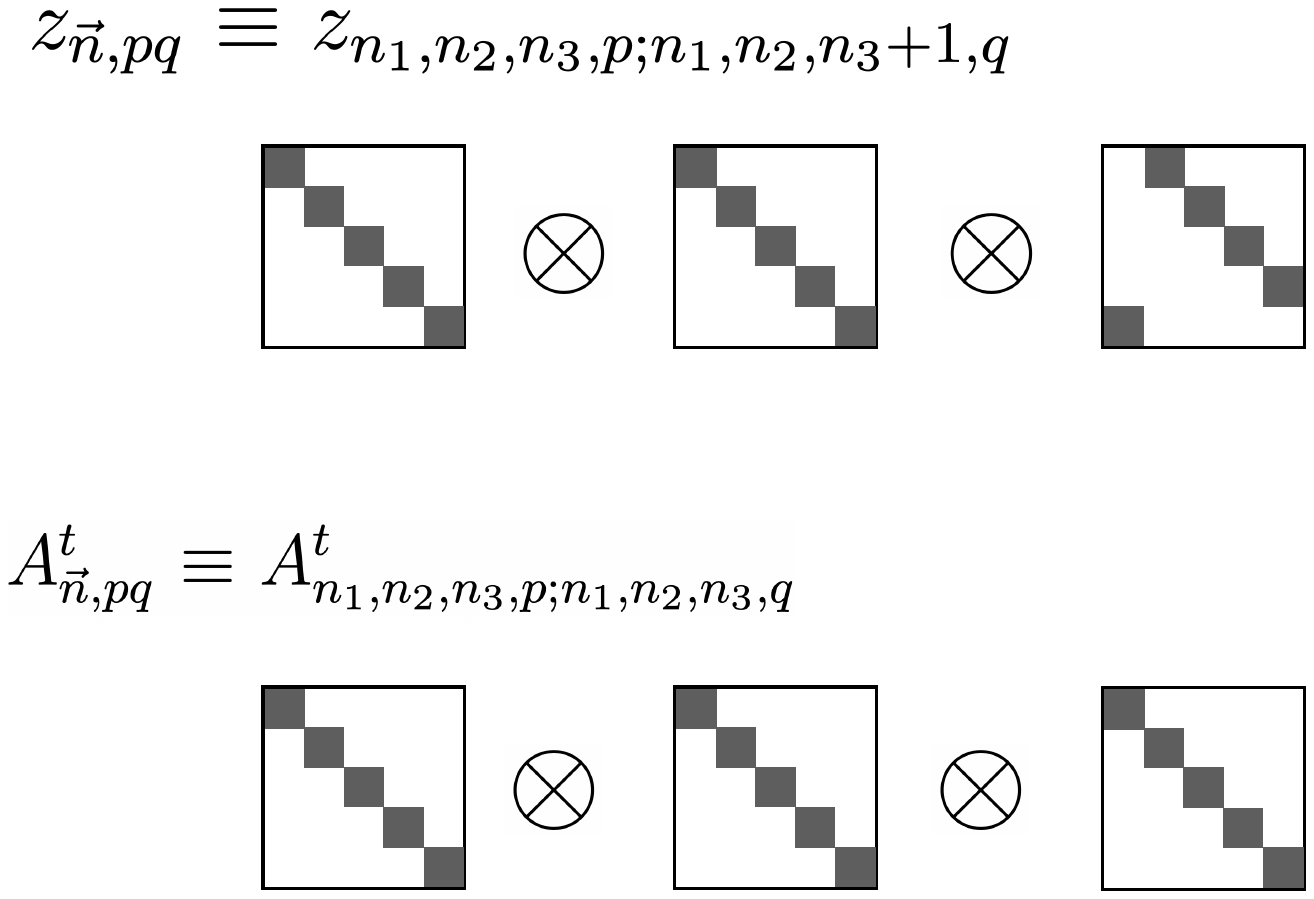}}
\end{center}
\caption{
A schematic picture of the embedding of the orbifold lattice into the matrix model, taken from Ref.~\cite{Buser:2020cvn}. 
Nonzero entries \eqref{orbifold_lattice_embedding} are shown in gray. 
}\label{fig:orbifold_lattice}
\end{figure}

The starting point is the Yang-Mills matrix model with $6$ scalar fields, which is sometimes called {\it mother theory}. The lattice theory obtained by applying orbifold projection on this theory is called {\it daughter theory}. 
The Lagrangian of the mother theory is 
\begin{align}
L_{\rm mother}
=
{\rm Tr}\left(
\frac{1}{2}\sum_{I}(D_tX_I)^2
+
\frac{g_{\rm 1d}^2}{4}\sum_{I,J}[X_I,X_J]^2
\right)\, . 
\end{align}
Here, the matrices $X_{I=1,\cdots,6}$ are $N_{\rm mat}\times N_{\rm mat}$ and Hermitian, and $D_t$ is the covariant derivative that acts on $X_I$
\begin{align}
D_tX_I
=
\partial_tX_I
+
ig_{\rm 1d}[A_t,X_I]\, . 
\end{align}
The coupling constant $g_{\rm 1d}$ is related to $g_{\rm 4d}$ by $g^2_{\rm 4d}=a^3g^2_{\rm 1d}$. 
To obtain the daughter theory with a U($N$) gauge group and lattice volume $L^3$, we take the matrix size $N_{\rm mat}$ to be $N_{\rm mat}=NL^3$. 
We introduce complex matrices $x$, $y$ and $z$ as
\begin{align}
x
=
\frac{X_1+iX_2}{\sqrt{2}}\, , 
\qquad
y
=
\frac{X_3+iX_4}{\sqrt{2}}\, , 
\qquad
z
=
\frac{X_5+iX_6}{\sqrt{2}}\, . 
\label{complex-vs-Hermitian}
\end{align}
Using the notation $\bar{x}=x^\dagger$, $\bar{y}=y^\dagger$ and $\bar{z}=z^\dagger$, the Lagrangian can be written as
\begin{align}
L_{\rm mother}
&=
{\rm Tr}\Biggl(
|D_tx|^2
+
|D_ty|^2
+
|D_tz|^2
-
\frac{g_{\rm 1d}^2}{2}\left|
[x,\bar{x}]
+
[y,\bar{y}]
+
[z,\bar{z}]
\right|^2
\nonumber\\
& 
\qquad
-
2g_{\rm 1d}^2
\left(
|[x,y]|^2
+
|[y,z]|^2
+
|[z,x]|^2
\right)
\Biggl)\, . 
\end{align}
To define the orbifold projection, we introduce the \textit{clock} matrices 
\begin{align}
C_1
&=
\Omega
\otimes
\textbf{1}_L
\otimes
\textbf{1}_L
\otimes
\textbf{1}_N\, , 
\nonumber\\
C_2
&=
\textbf{1}_L
\otimes
\Omega
\otimes
\textbf{1}_L
\otimes
\textbf{1}_N\, ,
\nonumber\\
C_3
&=
\textbf{1}_L
\otimes
\textbf{1}_L
\otimes
\Omega
\otimes
\textbf{1}_N\, , 
\end{align}
where
\begin{align}
\Omega
=
{\rm diag}
\left(
1,\omega, \omega^2,\cdots,\omega^{L-1}
\right)\, , 
\qquad
\omega
=
e^{-2\pi i/L}\, . 
\end{align}
By using the clock matrices, the orbifold projection condition can be written as 
\begin{align}
C_ixC_i^{-1}
=
\omega^{r_{x,i}}x\, , 
\qquad
C_iyC_i^{-1}
=
\omega^{r_{y,i}}y\, , 
\qquad
C_izC_i^{-1}
=
\omega^{r_{z,i}}z\, ,  
\qquad
C_iA_tC_i^{-1}
=
\omega^{r_{A,i}}A_t\, ,  
\end{align}
where
\begin{align}
\vec{r}_x
=
(1,0,0)\, , 
\qquad
\vec{r}_y
=
(0,1,0)\, , 
\qquad
\vec{r}_z
=
(0,0,1)\, ,
\qquad
\vec{r}_A
=
(0,0,0)\, . 
\end{align}
To label the matrix entries, we use $n_{1,2,3},n'_{1,2,3}=1,2,\cdots,L$ and $p,q=1,2,\cdots,N$ instead of $i,j=1,2,\cdots,N_{\rm mat}=NL^3$ 
respecting the tensor structure of the clock matrices, and we introduce the following convention:
\begin{align}
x_{ij}
&= 
x_{n_1,n_2,n_3,p;n'_1,n'_2,n'_3,q}\, ,
\nonumber\\
i
&=
p+(n_1-1)N+(n_2-1)NL+(n_3-1)NL^2\, ,
\nonumber\\ 
j
&=
q+(n'_1-1)N+(n'_2-1)NL+(n'_3-1)NL^2\, .
\end{align}
Then, the only entries that survive the orbifold projection are
\begin{align}
Z_{1,\vec{n},pq}
&\equiv
x_{\vec{n},pq}
\equiv
x_{n_1,n_2,n_3,p;n_1+1,n_2,n_3,q}\, , 
\nonumber\\
Z_{2,\vec{n},pq}
&\equiv
y_{\vec{n},pq}
\equiv
y_{n_1,n_2,n_3,p;n_1,n_2+1,n_3,q}\, , 
\nonumber\\
Z_{3,\vec{n},pq}
&\equiv
z_{\vec{n},pq}
\equiv
z_{n_1,n_2,n_3,p;n_1,n_2,n_3+1,q}\, , 
\nonumber\\
A_{t,\vec{n},pq}
&\equiv
A_{t,n_1,n_2,n_3,p;n_1,n_2,n_3,q}\, .
\label{orbifold_lattice_embedding}
\end{align}
Here periodic boundary conditions are assumed for $n_1,n_2$ and $n_3$. 
These nonzero entries are visualized in Fig.~\ref{fig:orbifold_lattice}. 
We interpret $\vec{n}$ as spatial lattice points. Then, the $Z_{j,\vec{n}}$ are interpreted as complex link variables.  
In this way we obtain the lattice Lagrangian \eqref{eq:lattice-action}. 

Now we include the fermions. We introduce fermions in the fundamental representation to the matrix model, as 
\begin{align}
L_{\rm mother,naive}
=
a^3\left\{
-
i\psi^\dagger D_t\psi
+
\frac{i}{2a}\sqrt{\frac{2g_{\rm 4d}^2}{a}}
\sum_{j=1}^3
\left(
\bar{\psi}\gamma^j x_{j}\psi
-
\bar{\psi}\gamma^j x^\dagger_{j}\psi
\right)
+
m\bar{\psi}\psi
\right\}
\end{align}
and
\begin{align}
L_{\rm mother,Wilson}
=
a^3\cdot\frac{i}{2a}\sum_{j=1}^3
\left(\bar{\psi}-\sqrt{\frac{2g_{\rm 4d}^2}{a}}\bar{\psi}\bar{x}_{j}\right)
\left(\psi-\sqrt{\frac{2g_{\rm 4d}^2}{a}}x_{j}\psi\right)\, . 
\end{align}
Here $x_{j=1,2,3}$ stands for $x,y$, and $z$. 
Then, we do not do anything to fermions and keep all $N_{\rm mat}=NL^3$ components untouched, while projecting $x_{j=1,2,3}$ and $A_t$ as we did above. Thus, we obtain $L_{\rm mother, naive}$ and $L_{\rm Wilson}$ defined by \eqref{naive_fermion_spatial_lattice} and \eqref{Wilson_fermion_spatial_lattice}.

\section{Implementation of \texorpdfstring{$\hat{p}^2$}{p²} by NOT, CNOT and CCNOT gates and with auxiliary qubits}\label{sec:plus_1}
In this appendix, we discuss how $\ket{n+1}\bra{n}$ in $\hat{p}^2$ can be realized by replacing logic gates on a classical computer by corresponding quantum gates.

Let us recall how the sum of two non-negative integers is coded by logic gates.
The simplest case is addition of $x=0$ or 1 and $x'=0$ or 1. Let $x+x'\equiv (y_1,y_2)=2y_2+y_1$. Then, we have $y_1=\mathrm{XOR}(x,x')$ and `carry up' $y_2=\mathrm{AND}(x,x')$: 
\begin{align}
\begin{array}{|c|c||c|c|}
\hline
x & x' & y_1=\mathrm{XOR}(x,x') & y_2=\mathrm{AND}(x,x')\\
\hline
\hline
0 & 0 & 0 & 0\\
\hline
1 & 0 & 1 & 0\\
\hline
0 & 1 & 1 & 0\\
\hline
1 & 1 & 0 & 1\\
\hline
\end{array}
 \end{align}
Next, let us consider addition of two-bit numbers $(x_1,x_2)\equiv 2x_2+x_1$ and $(x'_1,x'_2)\equiv 2x'_2+x'_1$. 
Let the answer be $(y_1,y_2,y_3)\equiv 2^2 y_3+2y_2+y_1$. Obviously, $y_1=\mathrm{XOR}(x_1,x'_1)$. To get $y_2$ and $y_3$, we need to add three numbers: $x_2, x'_2$ and carry up, $x''_2=\mathrm{AND}(x_1,x'_1)$. We can easily write down the input-output table:
\begin{align}
\begin{array}{|c|c|c||c|c|}
\hline
x_2 & x'_2 & x''_2 & y_2 & y_3\\
\hline
\hline
0 & 0 & 0 & 0 & 0\\
\hline
1 & 0 & 0 & 1 & 0\\
\hline
0 & 1 & 0 & 1 & 0\\
\hline
0 & 0 & 1 & 1 & 0\\
\hline
1 & 1 & 0 & 0 & 1\\
\hline
1 & 0 & 1 & 0 & 1\\
\hline
0 & 1 & 1 & 0 & 1\\
\hline
1 & 1 & 1 & 1 & 1\\
\hline
\end{array}
 \end{align} 
Unfortunately, however, there is no simple pattern. So, let us break the sum into three steps. 
Firstly, we compute $x_2+x'_2=(z_2,z_3)$. Then, we calculate $z_2+x''_2=(w_2,w_3)$. Then, by construction, 
$y_2=w_2$ and $y_3=z_3+w_3$. In this case, $z_3$ and $w_3$ cannot be 1 simultaneously, and hence, $y_3=z_3+w_3=\mathrm{XOR}(z_3,w_3)=\mathrm{OR}(z_3,w_3)$. 
{\tiny
\begin{align}
\begin{array}{|c|c|c||c|c||c|c||c|c|}
\hline
x_2 & x'_2 & x''_2& z_2=\mathrm{XOR}(x_2,x'_2) & z_3=\mathrm{AND}(x_2,x'_2) & w_2=\mathrm{XOR}(z_2,x''_2) & w_3=\mathrm{AND}(z_2,x''_2) & y_2=w_2 & y_3=z_3+w_3\\
\hline
\hline
0 & 0 & 0 & 0 & 0 & 0 & 0 & 0 & 0\\
\hline
1 & 0 & 0 & 1 & 0 & 1 & 0 & 1 & 0\\
\hline
0 & 1 & 0 & 1 & 0 & 1 & 0 & 1 & 0\\
\hline
0 & 0 & 1 & 0 & 0 & 1 & 0 & 1 & 0\\
\hline
1 & 1 & 0 & 0 & 1 & 0 & 0 & 0 & 1\\
\hline
1 & 0 & 1 & 1 & 0 & 0 & 1 & 0 & 1\\
\hline
0 & 1 & 1 & 1 & 0 & 0 & 1 & 0 & 1\\
\hline
1 & 1 & 1 & 0 & 1 & 1 & 0 & 1 & 1\\
\hline
\end{array}
 \end{align} 
 }
 The generalization to larger numbers with more digits is straightforward.  
We can use exactly the same circuit on quantum devices, just replacing the classical gates with quantum gates. 

To describe the momentum part of the Hamiltonian, we need only $n\to n\pm 1$. Therefore, we can simplify the construction. To code $n\to n+1$, we only need $x'_1=1$ and $x'_2=x'_3=\cdots = 0$ in the above. The tables simplify to 
\begin{align}
\begin{array}{|c|c||c|c|}
\hline
x_1 & x'_1 & y_1=\mathrm{XOR}(x_1,x'_1) & x''_2=\mathrm{AND}(x_1,x'_1)\\
\hline
\hline
0 & 1 & 1 & 0\\
\hline
1 & 1 & 0 & 1\\
\hline
\end{array}
 \end{align}
 and, for $j=2$, 
\begin{align}
\begin{array}{|c|c|c||c|c|}
\hline
x_j & x'_j & x''_j & y_j = \mathrm{XOR}(x_j,x''_j) & x''_{j+1}=\mathrm{AND}(x_j,x''_j)\\
\hline
\hline
0 & 0 & 0 & 0 & 0\\
\hline
1 & 0 & 0 & 1 & 0\\
\hline
0 & 0 & 1 & 1 & 0\\
\hline
1 & 0 & 1 & 0 & 1\\
\hline
\end{array}
\end{align}
Note that $y_1=\mathrm{XOR}(x_1,x'_1) = \mathrm{NOT}(x_1)$ and $x''_2=\mathrm{AND}(x_1,x'_1)=x_1$ when $x'_1=1$. 
  
To realize it, we can use $K-1$ auxiliary qubits set to $\ket{0}$ (which can be reduced to two qubits if we reuse the same qubits after resetting them), one NOT (=$\hat{\sigma}_x$), $K$ CNOT, and $K-2$ CCNOT gates (also known as Toffoli gates). 
CNOT is defined by 
\begin{align}
\mathrm{CNOT}:\quad
\ket{a}\ket{b}\to\ket{a}\cdot(\hat{\sigma}_x)^a\ket{b}\, ,
\end{align}
or equivalently, 
\begin{align}
\mathrm{CNOT}:\quad
\ket{0}\ket{0}\to\ket{0}\ket{0}\, ,
\quad
\ket{0}\ket{1}\to\ket{0}\ket{1}\, ,
\quad
\ket{1}\ket{0}\to\ket{0}\ket{1}\, ,
\quad
\ket{1}\ket{1}\to\ket{1}\ket{0}\, .
\end{align}
CCNOT is defined by 
\begin{align}
\mathrm{CCNOT}:\quad
\ket{a}\ket{b}\ket{c}
\to
\ket{a}\ket{b}\cdot(\hat{\sigma}_x)^{ab}\ket{c}\, .
\end{align}
Namely, the last qubit $\ket{c}$ flips if $\ket{a}=\ket{b}=\ket{1}$. 
With all of these, we can  construct $\ket{n}\to\ket{n+1}$ as follows (underbar indicates the target qubit):
\begin{align}
&
\textcolor{red}{\ket{x_1}}\ket{x_2}\ket{x_3}\ket{x_4}\cdots\ket{x_K}
\textcolor{red}{\underline{\ket{0}}}\ket{0}\ket{0}\ket{0}\cdots\ket{0}\quad\leftarrow\mathrm{CNOT}
\nonumber\\
&
\textcolor{red}{\ket{x_1}}\ket{x_2}\ket{x_3}\ket{x_4}\cdots\ket{x_K}
\textcolor{red}{\ket{x''_2}}\ket{0}\ket{0}\ket{0}\cdots\ket{0}
\nonumber\\
&
\textcolor{blue}{\underline{\ket{x_1}}}\ket{x_2}\ket{x_3}\ket{x_4}\cdots\ket{x_K}
\ket{x''_2}\ket{0}\ket{0}\ket{0}\cdots\ket{0}\quad\leftarrow\mathrm{NOT}=\hat{\sigma}_x
\nonumber\\
&
\textcolor{blue}{\ket{y_1}}\ket{x_2}\ket{x_3}\ket{x_4}\cdots\ket{x_K}
\ket{x''_2}\ket{0}\ket{0}\ket{0}\cdots\ket{0}
\nonumber\\
&
\ket{y_1}\textcolor{red}{\ket{x_2}}\ket{x_3}\ket{x_4}\cdots\ket{x_K}
\textcolor{red}{\ket{x''_2}\underline{\ket{0}}}\ket{0}\ket{0}\cdots\ket{0}\quad\leftarrow\mathrm{CCNOT}
\nonumber\\
&
\ket{y_1}\textcolor{red}{\ket{x_2}}\ket{x_3}\ket{x_4}\cdots\ket{x_K}
\textcolor{red}{\ket{x''_2}\ket{x''_3}}\ket{0}\ket{0}\cdots\ket{0}
\nonumber\\
&
\ket{y_1}\textcolor{blue}{\underline{\ket{x_2}}}\ket{x_3}\ket{x_4}\cdots\ket{x_K}
\textcolor{blue}{\ket{x''_2}}\ket{x''_3}\ket{0}\cdots\ket{0}\quad\leftarrow\mathrm{CNOT}
\nonumber\\
&
\ket{y_1}\textcolor{blue}{\ket{y_2}}\ket{x_3}\ket{x_4}\cdots\ket{x_K}
\textcolor{blue}{\ket{x''_2}}\ket{x''_3}\ket{0}\cdots\ket{0}
\nonumber\\
&
\ket{y_1}\ket{y_2}\textcolor{red}{\ket{x_3}}\ket{x_4}\cdots\ket{x_K}
\ket{x''_2}\textcolor{red}{\ket{x''_3}\underline{\ket{0}}}\cdots\ket{0}\quad\leftarrow\mathrm{CCNOT}
\nonumber\\
&
\ket{y_1}\ket{y_2}\textcolor{red}{\ket{x_3}}\ket{x_4}\cdots\ket{x_K}
\ket{x''_2}\textcolor{red}{\ket{x''_3}\ket{x''_4}}\ket{0}\cdots\ket{0}
\nonumber\\
&
\ket{y_1}\ket{y_2}\textcolor{blue}{\underline{\ket{x_3}}}\ket{x_4}\cdots\ket{x_K}
\ket{x''_2}\textcolor{blue}{\ket{x''_3}}\ket{x''_4}\cdots\ket{0}\quad\leftarrow\mathrm{CNOT}
\nonumber\\
&
\ket{y_1}\ket{y_2}\textcolor{blue}{\ket{y_3}}\ket{x_4}\cdots\ket{x_K}
\ket{x''_2}\textcolor{blue}{\ket{x''_3}}\ket{x''_4}\cdots\ket{0}
\nonumber\\
&\cdots\cdots
\nonumber\\
&\cdots\cdots
\nonumber\\
&
\ket{y_1}\ket{y_2}\ket{y_3}\cdots\textcolor{red}{\ket{x_{K-1}}}\ket{x_K}
\ket{x''_2}\ket{x''_3}\cdots\textcolor{red}{\ket{x''_{K-1}}\underline{\ket{0}}}\quad\leftarrow\mathrm{CCNOT} 
\nonumber\\
&
\ket{y_1}\ket{y_2}\ket{y_3}\cdots\textcolor{red}{\ket{x_{K-1}}}\ket{x_K}
\ket{x''_2}\ket{x''_3}\cdots\textcolor{red}{\ket{x''_{K-1}}\ket{x''_K}}
\nonumber\\
&
\ket{y_1}\ket{y_2}\ket{y_3}\cdots\textcolor{blue}{\underline{\ket{x_{K-1}}}}\ket{x_K}
\ket{x''_2}\ket{x''_3}\cdots\textcolor{blue}{\ket{x''_{K-1}}}\ket{x''_K}\quad\leftarrow\mathrm{CNOT}
\nonumber\\
&
\ket{y_1}\ket{y_2}\ket{y_3}\cdots\textcolor{blue}{\ket{y_{K-1}}}\ket{x_K}
\ket{x''_2}\ket{x''_3}\cdots\textcolor{blue}{\ket{x''_{K-1}}}\ket{x''_K}
\nonumber\\
&
\ket{y_1}\ket{y_2}\ket{y_3}\cdots\ket{y_{K-1}}\textcolor{red}{\underline{\ket{x_{K}}}}
\ket{x''_2}\ket{x''_3}\cdots\ket{x''_{K-1}}\textcolor{red}{\ket{x''_{K}}}\quad\leftarrow\mathrm{CNOT} 
\nonumber\\
&
\ket{y_1}\ket{y_2}\ket{y_3}\cdots\ket{y_{K-1}}\textcolor{red}{\ket{y_{K}}}
\ket{x''_2}\ket{x''_3}\cdots\ket{x''_{K-1}}\textcolor{red}{\ket{x''_{K}}}
\nonumber\\
&
\end{align}
In the last step, we did not use CCNOT to get  $\ket{x''_{K+1}}$ because of the periodic boundary condition $\ket{\Lambda}=\ket{0}$.

\bibliographystyle{utphys}
\bibliography{ref-orbifold-vs-KS}

\providecommand{\href}[2]{#2}\begingroup\raggedright\begin{thebibliography}{10}

\bibitem{Bauer:2022hpo}
C.~W. Bauer {\em et~al.}, ``{Quantum Simulation for High-Energy Physics},''
  \href{http://dx.doi.org/10.1103/PRXQuantum.4.027001}{{\em PRX Quantum}
  {\bfseries 4} no.~2, (2023) 027001},
  \href{http://arxiv.org/abs/2204.03381}{{\ttfamily arXiv:2204.03381
  [quant-ph]}}.

\bibitem{Bauer:2023qgm}
C.~W. Bauer, Z.~Davoudi, N.~Klco, and M.~J. Savage, ``{Quantum simulation of
  fundamental particles and forces},''
  \href{http://dx.doi.org/10.1038/s42254-023-00599-8}{{\em Nature Rev. Phys.}
  {\bfseries 5} no.~7, (2023) 420--432}.

\bibitem{Wiese:2013uua}
U.-J. Wiese, ``{Ultracold Quantum Gases and Lattice Systems: Quantum Simulation
  of Lattice Gauge Theories},''
  \href{http://dx.doi.org/10.1002/andp.201300104}{{\em Annalen Phys.}
  {\bfseries 525} (2013) 777--796},
  \href{http://arxiv.org/abs/1305.1602}{{\ttfamily arXiv:1305.1602
  [quant-ph]}}.

\bibitem{Zohar:2015hwa}
E.~Zohar, J.~I. Cirac, and B.~Reznik, ``{Quantum Simulations of Lattice Gauge
  Theories using Ultracold Atoms in Optical Lattices},''
  \href{http://dx.doi.org/10.1088/0034-4885/79/1/014401}{{\em Rept. Prog.
  Phys.} {\bfseries 79} no.~1, (2016) 014401},
  \href{http://arxiv.org/abs/1503.02312}{{\ttfamily arXiv:1503.02312
  [quant-ph]}}.

\bibitem{Dalmonte:2016alw}
M.~Dalmonte and S.~Montangero, ``{Lattice gauge theory simulations in the
  quantum information era},''
  \href{http://dx.doi.org/10.1080/00107514.2016.1151199}{{\em Contemp. Phys.}
  {\bfseries 57} no.~3, (2016) 388--412},
  \href{http://arxiv.org/abs/1602.03776}{{\ttfamily arXiv:1602.03776
  [cond-mat.quant-gas]}}.

\bibitem{Banuls:2019bmf}
M.~C. Ba\~nuls {\em et~al.}, ``{Simulating Lattice Gauge Theories within
  Quantum Technologies},''
  \href{http://dx.doi.org/10.1140/epjd/e2020-100571-8}{{\em Eur. Phys. J. D}
  {\bfseries 74} no.~8, (2020) 165},
  \href{http://arxiv.org/abs/1911.00003}{{\ttfamily arXiv:1911.00003
  [quant-ph]}}.

\bibitem{Aidelsburger:2021mia}
M.~Aidelsburger {\em et~al.}, ``{Cold atoms meet lattice gauge theory},''
  \href{http://dx.doi.org/10.1098/rsta.2021.0064}{{\em Phil. Trans. Roy. Soc.
  Lond. A} {\bfseries 380} (2021) 20210064},
  \href{http://arxiv.org/abs/2106.03063}{{\ttfamily arXiv:2106.03063
  [cond-mat.quant-gas]}}.

\bibitem{Zohar:2021nyc}
E.~Zohar, ``{Quantum simulation of lattice gauge theories in more than one
  space dimension\textemdash{}requirements, challenges and methods},''
  \href{http://dx.doi.org/10.1098/rsta.2021.0069}{{\em Phil. Trans. A. Math.
  Phys. Eng. Sci.} {\bfseries 380} no.~2216, (2021) 20210069},
  \href{http://arxiv.org/abs/2106.04609}{{\ttfamily arXiv:2106.04609
  [quant-ph]}}.

\bibitem{Klco:2021lap}
N.~Klco, A.~Roggero, and M.~J. Savage, ``{Standard model physics and the
  digital quantum revolution: thoughts about the interface},''
  \href{http://dx.doi.org/10.1088/1361-6633/ac58a4}{{\em Rept. Prog. Phys.}
  {\bfseries 85} no.~6, (2022) 064301},
  \href{http://arxiv.org/abs/2107.04769}{{\ttfamily arXiv:2107.04769
  [quant-ph]}}.

\bibitem{Kaplan:2002wv}
D.~B. Kaplan, E.~Katz, and M.~Unsal, ``{Supersymmetry on a spatial lattice},''
  \href{http://dx.doi.org/10.1088/1126-6708/2003/05/037}{{\em JHEP} {\bfseries
  05} (2003) 037}, \href{http://arxiv.org/abs/hep-lat/0206019}{{\ttfamily
  arXiv:hep-lat/0206019}}.

\bibitem{Cohen:2003xe}
A.~G. Cohen, D.~B. Kaplan, E.~Katz, and M.~Unsal, ``{Supersymmetry on a
  Euclidean space-time lattice. 1. A Target theory with four supercharges},''
  \href{http://dx.doi.org/10.1088/1126-6708/2003/08/024}{{\em JHEP} {\bfseries
  08} (2003) 024}, \href{http://arxiv.org/abs/hep-lat/0302017}{{\ttfamily
  arXiv:hep-lat/0302017}}.

\bibitem{Cohen:2003qw}
A.~G. Cohen, D.~B. Kaplan, E.~Katz, and M.~Unsal, ``{Supersymmetry on a
  Euclidean space-time lattice. 2. Target theories with eight supercharges},''
  \href{http://dx.doi.org/10.1088/1126-6708/2003/12/031}{{\em JHEP} {\bfseries
  12} (2003) 031}, \href{http://arxiv.org/abs/hep-lat/0307012}{{\ttfamily
  arXiv:hep-lat/0307012}}.

\bibitem{Kaplan:2005ta}
D.~B. Kaplan and M.~Unsal, ``{A Euclidean lattice construction of
  supersymmetric Yang-Mills theories with sixteen supercharges},''
  \href{http://dx.doi.org/10.1088/1126-6708/2005/09/042}{{\em JHEP} {\bfseries
  09} (2005) 042}, \href{http://arxiv.org/abs/hep-lat/0503039}{{\ttfamily
  arXiv:hep-lat/0503039}}.

\bibitem{Wilson:1974sk}
K.~G. Wilson, ``{Confinement of Quarks},''
  \href{http://dx.doi.org/10.1103/PhysRevD.10.2445}{{\em Phys. Rev. D}
  {\bfseries 10} (1974) 2445--2459}.

\bibitem{Kogut:1974ag}
J.~B. Kogut and L.~Susskind, ``{Hamiltonian Formulation of Wilson's Lattice
  Gauge Theories},'' \href{http://dx.doi.org/10.1103/PhysRevD.11.395}{{\em
  Phys. Rev. D} {\bfseries 11} (1975) 395--408}.

\bibitem{Zohar:2014qma}
E.~Zohar and M.~Burrello, ``{Formulation of lattice gauge theories for quantum
  simulations},'' \href{http://dx.doi.org/10.1103/PhysRevD.91.054506}{{\em
  Phys. Rev. D} {\bfseries 91} no.~5, (2015) 054506},
  \href{http://arxiv.org/abs/1409.3085}{{\ttfamily arXiv:1409.3085
  [quant-ph]}}.

\bibitem{Buser:2020cvn}
A.~J. Buser, H.~Gharibyan, M.~Hanada, M.~Honda, and J.~Liu, ``{Quantum
  simulation of gauge theory via orbifold lattice},''
  \href{http://dx.doi.org/10.1007/JHEP09(2021)034}{{\em JHEP} {\bfseries 09}
  (2021) 034}, \href{http://arxiv.org/abs/2011.06576}{{\ttfamily
  arXiv:2011.06576 [hep-th]}}.

\bibitem{Hayata:2023bgh}
T.~Hayata and Y.~Hidaka, ``{q deformed formulation of Hamiltonian SU(3)
  Yang-Mills theory},'' \href{http://dx.doi.org/10.1007/JHEP09(2023)123}{{\em
  JHEP} {\bfseries 09} (2023) 123},
  \href{http://arxiv.org/abs/2306.12324}{{\ttfamily arXiv:2306.12324
  [hep-lat]}}.

\bibitem{Gustafson:2022xdt}
E.~J. Gustafson, H.~Lamm, F.~Lovelace, and D.~Musk, ``{Primitive quantum gates
  for an SU(2) discrete subgroup: Binary tetrahedral},''
  \href{http://dx.doi.org/10.1103/PhysRevD.106.114501}{{\em Phys. Rev. D}
  {\bfseries 106} no.~11, (2022) 114501},
  \href{http://arxiv.org/abs/2208.12309}{{\ttfamily arXiv:2208.12309
  [quant-ph]}}.

\bibitem{Alexandru:2023qzd}
A.~Alexandru, P.~F. Bedaque, A.~Carosso, M.~J. Cervia, E.~M. Murairi, and
  A.~Sheng, ``{Fuzzy Gauge Theory for Quantum Computers},''
  \href{http://arxiv.org/abs/2308.05253}{{\ttfamily arXiv:2308.05253
  [hep-lat]}}.

\bibitem{Bhattacharya:2020gpm}
T.~Bhattacharya, A.~J. Buser, S.~Chandrasekharan, R.~Gupta, and H.~Singh,
  ``{Qubit regularization of asymptotic freedom},''
  \href{http://dx.doi.org/10.1103/PhysRevLett.126.172001}{{\em Phys. Rev.
  Lett.} {\bfseries 126} no.~17, (2021) 172001},
  \href{http://arxiv.org/abs/2012.02153}{{\ttfamily arXiv:2012.02153
  [hep-lat]}}.

\bibitem{Hanada:2010qg}
M.~Hanada and I.~Kanamori, ``{Absence of sign problem in two-dimensional N =
  (2,2) super Yang-Mills on lattice},''
  \href{http://dx.doi.org/10.1007/JHEP01(2011)058}{{\em JHEP} {\bfseries 01}
  (2011) 058}, \href{http://arxiv.org/abs/1010.2948}{{\ttfamily arXiv:1010.2948
  [hep-lat]}}.

\bibitem{Ishibashi:1996xs}
N.~Ishibashi, H.~Kawai, Y.~Kitazawa, and A.~Tsuchiya, ``{A Large N reduced
  model as superstring},''
  \href{http://dx.doi.org/10.1016/S0550-3213(97)00290-3}{{\em Nucl. Phys. B}
  {\bfseries 498} (1997) 467--491},
  \href{http://arxiv.org/abs/hep-th/9612115}{{\ttfamily arXiv:hep-th/9612115}}.

\bibitem{Hanada:2020uvt}
M.~Hanada, H.~Shimada, and N.~Wintergerst, ``{Color confinement and
  Bose-Einstein condensation},''
  \href{http://dx.doi.org/10.1007/JHEP08(2021)039}{{\em JHEP} {\bfseries 08}
  (2021) 039}, \href{http://arxiv.org/abs/2001.10459}{{\ttfamily
  arXiv:2001.10459 [hep-th]}}.

\bibitem{Rinaldi:2021jbg}
E.~Rinaldi, X.~Han, M.~Hassan, Y.~Feng, F.~Nori, M.~McGuigan, and M.~Hanada,
  ``{Matrix-Model Simulations Using Quantum Computing, Deep Learning, and
  Lattice Monte Carlo},''
  \href{http://dx.doi.org/10.1103/PRXQuantum.3.010324}{{\em PRX Quantum}
  {\bfseries 3} no.~1, (2022) 010324},
  \href{http://arxiv.org/abs/2108.02942}{{\ttfamily arXiv:2108.02942
  [quant-ph]}}.

\bibitem{Jordan:2012xnu}
S.~P. Jordan, K.~S.~M. Lee, and J.~Preskill, ``{Quantum Algorithms for Quantum
  Field Theories},'' \href{http://dx.doi.org/10.1126/science.1217069}{{\em
  Science} {\bfseries 336} (2012) 1130--1133},
  \href{http://arxiv.org/abs/1111.3633}{{\ttfamily arXiv:1111.3633
  [quant-ph]}}.

\bibitem{Jordan:2011ci}
S.~P. Jordan, K.~S.~M. Lee, and J.~Preskill, ``{Quantum Computation of
  Scattering in Scalar Quantum Field Theories},'' {\em Quant. Inf. Comput.}
  {\bfseries 14} (2014) 1014--1080,
  \href{http://arxiv.org/abs/1112.4833}{{\ttfamily arXiv:1112.4833 [hep-th]}}.

\bibitem{Klco:2018zqz}
N.~Klco and M.~J. Savage, ``{Digitization of scalar fields for quantum
  computing},'' \href{http://dx.doi.org/10.1103/PhysRevA.99.052335}{{\em Phys.
  Rev. A} {\bfseries 99} no.~5, (2019) 052335},
  \href{http://arxiv.org/abs/1808.10378}{{\ttfamily arXiv:1808.10378
  [quant-ph]}}.

\bibitem{Cowtan:2019loc}
A.~Cowtan, S.~Dilkes, R.~Duncan, W.~Simmons, and S.~Sivarajah, ``{Phase Gadget
  Synthesis for Shallow Circuits},''
  \href{http://dx.doi.org/10.4204/EPTCS.318.13}{{\em EPTCS} {\bfseries 318}
  (2020) 213--228}, \href{http://arxiv.org/abs/1906.01734}{{\ttfamily
  arXiv:1906.01734 [quant-ph]}}.

\bibitem{Sriluckshmy:2023leq}
P.~V. Sriluckshmy, V.~Pina-Canelles, M.~Ponce, M.~G. Algaba, F.~v. IV, and
  M.~Leib, ``{Optimal, hardware native decomposition of parameterized
  multi-qubit Pauli gates},''
  \href{http://dx.doi.org/10.1088/2058-9565/acfa20}{{\em Quantum Sci. Technol.}
  {\bfseries 8} no.~4, (2023) 045029},
  \href{http://arxiv.org/abs/2303.04498}{{\ttfamily arXiv:2303.04498
  [quant-ph]}}.

\bibitem{Algaba:2023enr}
M.~G. Algaba, P.~V. Sriluckshmy, M.~Leib, and F.~\v{S}imkovic, ``{Low-depth
  simulations of fermionic systems on square-grid quantum hardware},''
  \href{http://arxiv.org/abs/2302.01862}{{\ttfamily arXiv:2302.01862
  [quant-ph]}}.

\bibitem{Shaydulin:2023fpr}
R.~Shaydulin {\em et~al.}, ``{Evidence of Scaling Advantage for the Quantum
  Approximate Optimization Algorithm on a Classically Intractable Problem},''
  \href{http://arxiv.org/abs/2308.02342}{{\ttfamily arXiv:2308.02342
  [quant-ph]}}.

\bibitem{Farrell:2024fit}
R.~C. Farrell, M.~Illa, A.~N. Ciavarella, and M.~J. Savage, ``{Quantum
  Simulations of Hadron Dynamics in the Schwinger Model using 112 Qubits},''
  \href{http://arxiv.org/abs/2401.08044}{{\ttfamily arXiv:2401.08044
  [quant-ph]}}.

\bibitem{Alam:2021uuq}
{\bfseries SQMS} Collaboration, M.~S. Alam, S.~Hadfield, H.~Lamm, and A.~C.~Y.
  Li, ``{Primitive quantum gates for dihedral gauge theories},''
  \href{http://dx.doi.org/10.1103/PhysRevD.105.114501}{{\em Phys. Rev. D}
  {\bfseries 105} no.~11, (2022) 114501},
  \href{http://arxiv.org/abs/2108.13305}{{\ttfamily arXiv:2108.13305
  [quant-ph]}}.

\bibitem{Jakobs:2023lpp}
T.~Jakobs, M.~Garofalo, T.~Hartung, K.~Jansen, J.~Ostmeyer, D.~Rolfes,
  S.~Romiti, and C.~Urbach, ``{Canonical momenta in digitized Su(2) lattice
  gauge theory: definition and free theory},''
  \href{http://dx.doi.org/10.1140/epjc/s10052-023-11829-9}{{\em Eur. Phys. J.
  C} {\bfseries 83} no.~7, (2023) 669},
  \href{http://arxiv.org/abs/2304.02322}{{\ttfamily arXiv:2304.02322
  [hep-lat]}}.

\bibitem{Garofalo:2023zkd}
M.~Garofalo, T.~Hartung, T.~Jakobs, K.~Jansen, J.~Ostmeyer, D.~Rolfes,
  S.~Romiti, and C.~Urbach, ``{Testing the $\mathrm{SU}(2)$ lattice Hamiltonian
  built from $S_3$ partitionings},''
  \href{http://arxiv.org/abs/2311.15926}{{\ttfamily arXiv:2311.15926
  [hep-lat]}}.

\bibitem{Romiti:2023hbd}
S.~Romiti and C.~Urbach, ``{Digitizing lattice gauge theories in the magnetic
  basis: reducing the breaking of the fundamental commutation relations},''
  \href{http://arxiv.org/abs/2311.11928}{{\ttfamily arXiv:2311.11928
  [hep-lat]}}.

\bibitem{DAndrea:2023qnr}
I.~D'Andrea, C.~W. Bauer, D.~M. Grabowska, and M.~Freytsis, ``{New basis for
  Hamiltonian SU(2) simulations},''
  \href{http://dx.doi.org/10.1103/PhysRevD.109.074501}{{\em Phys. Rev. D}
  {\bfseries 109} no.~7, (2024) 074501},
  \href{http://arxiv.org/abs/2307.11829}{{\ttfamily arXiv:2307.11829
  [hep-ph]}}.

\bibitem{Hanada:2022pps}
M.~Hanada, J.~Liu, E.~Rinaldi, and M.~Tezuka, ``{Estimating truncation effects
  of quantum bosonic systems using sampling algorithms},''
  \href{http://dx.doi.org/10.1088/2632-2153/ad035c}{{\em Mach. Learn. Sci.
  Tech.} {\bfseries 4} no.~4, (2023) 045021},
  \href{http://arxiv.org/abs/2212.08546}{{\ttfamily arXiv:2212.08546
  [quant-ph]}}.

\bibitem{Sachdev:2015efa}
S.~Sachdev, ``{Bekenstein-Hawking Entropy and Strange Metals},''
  \href{http://dx.doi.org/10.1103/PhysRevX.5.041025}{{\em Phys. Rev. X}
  {\bfseries 5} no.~4, (2015) 041025},
  \href{http://arxiv.org/abs/1506.05111}{{\ttfamily arXiv:1506.05111
  [hep-th]}}.

\bibitem{Maldacena:2016hyu}
J.~Maldacena and D.~Stanford, ``{Remarks on the Sachdev-Ye-Kitaev model},''
  \href{http://dx.doi.org/10.1103/PhysRevD.94.106002}{{\em Phys. Rev. D}
  {\bfseries 94} no.~10, (2016) 106002},
  \href{http://arxiv.org/abs/1604.07818}{{\ttfamily arXiv:1604.07818
  [hep-th]}}.

\bibitem{Hanada:2023rkf}
M.~Hanada, A.~Jevicki, X.~Liu, E.~Rinaldi, and M.~Tezuka, ``{A model of
  randomly-coupled Pauli spins},''
  \href{http://arxiv.org/abs/2309.15349}{{\ttfamily arXiv:2309.15349
  [hep-th]}}.

\bibitem{Swingle:2023nvv}
B.~Swingle and M.~Winer, ``{A Bosonic Model of Quantum Holography},''
  \href{http://arxiv.org/abs/2311.01516}{{\ttfamily arXiv:2311.01516
  [hep-th]}}.

\bibitem{Danshita:2016xbo}
I.~Danshita, M.~Hanada, and M.~Tezuka, ``{Creating and probing the
  Sachdev-Ye-Kitaev model with ultracold gases: Towards experimental studies of
  quantum gravity},'' \href{http://dx.doi.org/10.1093/ptep/ptx108}{{\em PTEP}
  {\bfseries 2017} no.~8, (2017) 083I01},
  \href{http://arxiv.org/abs/1606.02454}{{\ttfamily arXiv:1606.02454
  [cond-mat.quant-gas]}}.

\bibitem{Brown:2019hmk}
A.~R. Brown, H.~Gharibyan, S.~Leichenauer, H.~W. Lin, S.~Nezami, G.~Salton,
  L.~Susskind, B.~Swingle, and M.~Walter, ``{Quantum Gravity in the Lab. I.
  Teleportation by Size and Traversable Wormholes},''
  \href{http://dx.doi.org/10.1103/PRXQuantum.4.010320}{{\em PRX Quantum}
  {\bfseries 4} no.~1, (2023) 010320},
  \href{http://arxiv.org/abs/1911.06314}{{\ttfamily arXiv:1911.06314
  [quant-ph]}}.

\bibitem{Gharibyan:2020bab}
H.~Gharibyan, M.~Hanada, M.~Honda, and J.~Liu, ``{Toward simulating
  superstring/M-theory on a quantum computer},''
  \href{http://dx.doi.org/10.1007/JHEP07(2021)140}{{\em JHEP} {\bfseries 07}
  (2021) 140}, \href{http://arxiv.org/abs/2011.06573}{{\ttfamily
  arXiv:2011.06573 [hep-th]}}.

\bibitem{Maldacena:2023acv}
J.~Maldacena, ``{A simple quantum system that describes a black hole},''
  \href{http://arxiv.org/abs/2303.11534}{{\ttfamily arXiv:2303.11534
  [hep-th]}}.

\bibitem{KU-private-communication}
D.~B. Kaplan and M.~Unsal, ``{Privare communication},''.

\end{thebibliography}\endgroup

\end{document}